\documentstyle[12pt,aaspp4]{article}
\lefthead{Bureau \& Carignan}
\righthead{Environment, Ram Pressure, and Shell Formation in HoII}
\slugcomment{Submitted to The Astronomical Journal}
\begin{document}
%
%
\newcommand{\kms}{km~s$^{-1}$}
\newcommand{\msun}{$M_{\mbox{\scriptsize \sun}}$}
\newcommand{\coltwo}[2]{\mbox{$\displaystyle\mbox{#1}\atop{\displaystyle\mbox{#2}}$}}
%
%
\title{Environment, Ram Pressure, and Shell Formation in HoII}
\author{M.\ Bureau\altaffilmark{1}\altaffilmark{2}}
\affil{Sterrewacht Leiden, Postbus~9513, 2300~RA Leiden, The Netherlands;
  bureau@strw.leidenuniv.nl}
\altaffiltext{1}{Hubble Fellow}
\altaffiltext{2}{Now at: Columbia Astrophysics Laboratory, 550~West
  120th~Street, 1027 Pupin Hall, Mail Code~5247, New~York, NY~10027,
  U.S.A.}
\and
\author{C.\ Carignan}
\affil{D\'{e}partement de Physique and Observatoire du Mont M\'{e}gantic,
  Universit\'{e} de Montr\'{e}al, C.~P.\ 6128, Succ.\ ``centre-ville'',
  Montr\'{e}al, Qu\'{e}bec, Canada H3C~3J7; carignan@astro.umontreal.ca}
\begin{abstract}
  Neutral hydrogen VLA D-array observations of the dwarf irregular galaxy
  HoII, a prototype galaxy for studies of shell formation, are presented.
  These were extracted from the multi-configuration dataset of Puche et al.\ 
  (\markcite{PWBR92}1992). \ion{H}{1} is detected to radii over 16\arcmin\ or
  $4R_{25}$, almost a factor of two better than previous studies. The total
  \ion{H}{1} mass $M_{\mbox{\scriptsize HI}}=6.44\times10^8$~\msun. The
  integrated \ion{H}{1} map has a comet-like appearance, with a large but
  faint component extending to the northwest and the \ion{H}{1} appearing
  compressed on the opposite side. This suggests that HoII is affected by ram
  pressure from an intragroup medium (IGM). The velocity field shows a clear
  rotating disk pattern and a rotation curve corrected for asymmetric drift
  was derived. However, the gas at large radii may not be in equilibrium.
  Puche et al.\ (\markcite{PWBR92}1992) multi-configuration data were also
  reanalyzed and it is shown that they overestimated their fluxes by over
  20\%.
  
  The rotation curve derived for HoII is well defined for $r\lesssim10$~kpc.
  For $10\lesssim r\lesssim18$~kpc, however, velocities are only defined on
  the approaching side, such that this part of the rotation curve should be
  used with caution.  An analysis of the mass distribution, using the whole
  extent of this rotation curve, yields a total mass of $6.3\times10^9$~\msun,
  of which $\approx80$\% is dark. Similarly to what is seen in many dwarfs,
  there is more luminous mass in \ion{H}{1} than in stars. One peculiarity,
  however, is that luminous matter dominates within the optical body of the
  galaxy and dark matter only in the outer parts, analogous to what is seen in
  massive spirals rather than dwarfs.
  
  HoII lies northeast of the M81 group's core, along with Kar~52 (M81~Dwarf~A)
  and UGC~4483. No signs of interaction are observed, however, and it is
  argued that HoII is part of the NGC~2403 subgroup, infalling towards M81. A
  case is made for ram pressure stripping and an IGM in the M81 group.
  Stripping of the outer parts of the disk would require an IGM density
  $n_{\mbox{\tiny IGM}}\gtrsim4.0\times10^{-6}$~atoms~cm$^{-3}$ at the
  location of HoII. This corresponds to $\sim$1\% of the virial mass of the
  group uniformly distributed over a volume just enclosing HoII and it is
  consistent with the known X-ray properties of small groups. The \ion{H}{1}
  tail is consistent with additional turbulent viscous stripping and
  evaporation, at least for low IGM temperatures.
  
  It is argued that existing observations of HoII do not support
  self-propagating star formation scenarios, whereby the \ion{H}{1} holes and
  shells are created by supernova explosions and stellar winds. Many
  \ion{H}{1} holes are located in low surface density regions of the disk,
  where no star formation is expected or observed. Alternative mechanisms are
  discussed and it is suggested that ram pressure can help. Ram pressure has
  the capacity to enlarge preexisting holes and lower their creation energies,
  helping to bridge the gap between the observed star formation rate and that
  required to create the holes.
\end{abstract}
\section{Introduction\label{sec:introduction}}
\nopagebreak
Spiral galaxies have a complex interstellar medium (ISM) pierced by numerous
cavities. These were first identified in the Galaxy (Heiles
\markcite{h79}1979, \markcite{h84}1984) and M31 (Brinks \markcite{b81}1981;
Brinks \& Bajaja \markcite{bb86}1986), but similar structures were quickly
discovered in other nearby galaxies (e.g.\ Deul \& den Hartog
\markcite{dh90}1990 for M33; Kamphuis \markcite{k93}1993 for M101 and
NGC~6946). The majority of these cavities are generally thought to arise from
the combined effects of supernova explosions (SNe) and stellar winds (see
Tenorio-Tagle \& Bodenheimer \markcite{tb88}1988 and van der Hulst
\markcite{h96}1996 for reviews), consistent with a three-phase picture of the
ISM (Cox \& Smith \markcite{cs74}1974; McKee \& Ostriker \markcite{mo77}1977).
In late-type spirals and dwarfs, cavities are long-lived due to a combination
of low shear, a shallow gravitational potential, the absence of spiral density
waves, and large scaleheights.

One of the best studied dwarf irregulars is without doubt HoII, in the nearby
M81 group of galaxies. Its basic properties are listed in
Table~\ref{ta:basic}. Puche et al.\ (\markcite{PWBR92}1992; hereafter
\markcite{PWBR92}PWBR92) cataloged and characterized over 50 \ion{H}{1} holes
and shells in HoII using multi-configuration observations from NRAO's
Very Large Array\footnote{The National Radio Astronomy Observatory is
a facility of the National Science Foundation operated under a
cooperative agreement by Associated Universities, Inc.} (VLA), arguing for
sequential star formation events in the disk. This picture has since been
criticized (e.g.\ Rhode et al.\ \markcite{rswr99}1999), but it was already
clear then that some holes required extremely high creation energies (up to
$2\times10^{53}$~ergs). \markcite{PWBR92}PWBR92 pointed out that HoII probably
contains some dark matter, but they commented only briefly on its large-scale
mass distribution, although it may be crucial to understand the evolution of
the shells and constrain the amount of gas expelled from the galaxy (e.g.\ 
Silich et al.\ \markcite{sfpt96}1996; Efstathiou \markcite{e00}2000).
\markcite{PWBR92}PWBR92 detected a small \ion{H}{1} cloud to the southwest,
pointing to a significant amount of \ion{H}{1} at large radii. This is
essential to constrain the dark matter distribution, but also hints at a
disturbed large-scale ISM or intergalactic medium (IGM). This is vindicated in
our study, which shows that HoII has an extended comet-like \ion{H}{1}
morphology, possibly caused by ram pressure.

\placetable{ta:basic}

In this paper, we will focus on HoII's large-scale \ion{H}{1} distribution,
looking at how it can help us understand its current properties and unravel
its evolutionary history. In \S~\ref{sec:data}, we reanalyze and discuss
\markcite{PWBR92}PWBR92's data, using exclusively the low-resolution D-array
observations. In \S~\ref{sec:mass}, we derive HoII's rotation curve and
discuss its mass distribution. The likelihood of ram pressure and the
properties of a putative IGM are discussed in \S~\ref{sec:ram+environment}. In
\S~\ref{sec:shells}, the creation of shells and supershells is discussed in
light of all available observations. We summarize our results and briefly
conclude in \S~\ref{sec:conclusions}.
\section{Data Reduction and Analysis\label{sec:data}}
\nopagebreak
The starting point for our analysis was the multi-configuration VLA
data of \markcite{PWBR92}PWBR92 (B, C, and D arrays; 1990 August 10,
1990 December 2, and 1991 March 6, respectively), which were
kindly made available to us by Puche. We only briefly review here the
calibration of the data and refer the reader to \markcite{PWBR92}PWBR92 for a
more extensive discussion.  The flux calibration and complex bandpass
corrections were provided by observations of 3C~286 in each configuration and
the amplitudes and phases were calibrated using 0859+681 and 0836+710, close
to the source. The data were also edited to remove bad data points and the
three $uv$ databases merged. The resulting $uv$ dataset was used as a starting
point for the present study and the subsequent analysis by
\markcite{PWBR92}PWBR92 has not been used. Table~\ref{ta:h1} summarizes the
observational set-up and global \ion{H}{1} properties as derived from our
work. All the data reduction was done using standard procedures in AIPS.

\placetable{ta:h1}
\subsection{Multi-Configuration Data\label{sec:multi}}
\nopagebreak
We first rereduced the multi-configuration data of \markcite{PWBR92}PWBR92 to
optimize the sensitivity of the observations and to take advantage of the
\ion{H}{1} cloud detected to the southwest of the galaxy to extend as far as
possible the kinematic information. The continuum was subtracted from the
merged $uv$ dataset using line-free channels and the AIPS task UVLIN, paying
special attention to a strong continuum source at $\alpha=8\mbox{h} 17\mbox{m}
05\mbox{s}$, $\delta=+70\arcdeg 52\arcmin 29\farcs8$ (B1950). Given the small
bandwidth used, HoII had no detectable continuum emission of its own (see,
however, Tongue \& Westpfahl \markcite{tw95}1995 for 90, 20, and 6~cm VLA
observations).

Various tests were then performed using both uniform weighting, natural
weighting, and various tapers and cleaning strategies. The application of a
taper decreases the resolution without increasing much the detected flux, so a
naturally weighted cube with no taper was produced (the cleaning was done with
the AIPS task MX). This results in a beam of $14\farcs8\times14\farcs5$ and a
rms noise per channel $\sigma=1.1$~mJy~beam$^{-1}$, essentially the same
as \markcite{PWBR92}PWBR92's medium resolution cube. This does not improve on
the work of \markcite{PWBR92}PWBR92, and the multi-configuration data will not
be discussed further in light of the goals stated in the introduction.
However, the fluxes published by \markcite{PWBR92}PWBR92 must be revised down,
due to complications when measuring fluxes of multi-configuration data. We
discuss this issue below.

During our tests, it was noticed that the total flux in the cleaned maps
decreased with increasing cleaning depth and/or increasing clean (restoring)
beam size. This could be traced to a very non-Gaussian beam.  Indeed, the beam
yielded by the multi-configuration data is strongly peaked, and its effective
area differs significantly from that of a Gaussian beam of equivalent
resolution, resulting in bad total fluxes when the residuals and restored
cleaned components of a map are summed (see H\"{o}gbom \markcite{h74}1974 for
a description of the CLEAN algorithm). The fluxes converge for very deep
cleaning, but this is impractical. J\"{o}rs\"{a}ter \& van Moorsel
\markcite{jv95}(1995) describe a simple corrective method to obtain reliable
fluxes in such a situation. We applied this method to our multi-configuration
naturally weighted data cube, channel by channel, to produce a clean (and
corrected) cube with a cleaning depth of $1.5\sigma$ and a restoring beam of
$15\arcsec\times15\arcsec$. The correction factor for the residuals in
individual channels varied from 0.20 to 0.40, with a mean of 0.30. An
identical cube without flux correction was also produced for comparison
purposes.

Both cubes were used to produce moment maps, using a Gaussian smoothing of 2
spatially and Hanning smoothing over 3 channels spectrally. For all practical
purposes, these moment maps are identical to those obtained by
\markcite{PWBR92}PWBR92 for their medium resolution data cube (see their
Fig.~3 and 4). The cubes were also used to derive global profiles (after
correcting for primary beam attenuation), summing the emission in each channel
using the moment~0 map as a mask. The corrected and uncorrected profiles are
shown in Fig.~\ref{fig:global_multi}.

\placefigure{fig:global_multi}

As expected, the shapes of the profiles are very similar, except at low fluxes
where the contribution from residuals is non-negligible. The derived systemic
velocities and velocity widths are identical and in agreement with those of
Tully \markcite{t88}(1988): $V_{\mbox{\scriptsize \sun}}=156\pm2$~\kms,
$\Delta V_{50}=57\pm3$~\kms, and $\Delta V_{20}=72\pm3$~\kms. These also compare
favorably with the global profile of \markcite{PWBR92}PWBR92's medium
resolution data cube, shown in their Fig.~2. On the other hand, the absolute
fluxes of the three profiles differ substantially. The fluxes in our corrected
cube are about 10\% lower than those of the uncorrected cube
($F_{\mbox{\scriptsize HI}}=266$ and 294~Jy~\kms\ respectively), in agreement
with a correction factor less than unity for the residuals. This was to be
expected since, before correction, the total flux decreased with increasing
cleaning depth, suggesting that the residuals were overestimated. Had we
cleaned less deep, the difference would have been even more important.
\markcite{PWBR92}PWBR92 report a total flux $F_{\mbox{\scriptsize
    HI}}=327$~Jy~\kms, 10\% higher than our {\em un}corrected cube. This can
be explained by the different restoring beams used (\markcite{PWBR92}PWBR92
used a $11\farcs0\times10\farcs9$ beam), that is, \markcite{PWBR92}PWBR92
overestimated their residuals with respect to ours by a factor approximately
equal to the ratio of the effective beam areas. Taking this into account, we
can reproduce \markcite{PWBR92}PWBR92's fluxes to within 1\%.

This leaves no doubt that the corrective procedure proposed by
J\"{o}rs\"{a}ter \& van Moorsel \markcite{jv95}(1995) must be used to obtain
reliable fluxes when dealing with complex beams, and that the fluxes derived
from our corrected cube must be preferred over those of
\markcite{PWBR92}PWBR92. Single-dish total fluxes reported in the literature
vary widely (Huchtmeier \& Richter \markcite{hr89}1989 report values from
$F_{\mbox{\scriptsize HI}}=164$ to 332~Jy~\kms). Furthermore, given the very
extended nature of the \ion{H}{1} in HoII (as shown in the next section), even
single-dish data could miss some emission, so it is difficult to estimate how
much flux the synthesis observations are missing.
\subsection{D-Array Data\label{sec:d}}
\nopagebreak
As stated previously, the main reason behind a reanalysis of
\markcite{PWBR92}PWBR92's data was to maximize the sensitivity to large-scale
structures. We thus created a new $uv$ dataset containing the D-array data
only, even though only 2 hours were spent observing in this configuration. The
discussion in the rest of this paper focuses mainly on these data.

We first subtracted the continuum emission from the D-array $uv$ dataset using
line-free channels and the AIPS task UVLIN, as for the multi-configuration
data. The emission extends approximately from $v_{\mbox{\scriptsize
    \sun}}=105$ to 215~\kms. A naturally weighted cube with no taper was then
created using the AIPS task MX. The resulting beam is
$66\farcs7\times46\farcs4$ and the rms noise in the line-free channels
$\sigma=2.75$~mJy~beam$^{-1}$. A cube cleaned to a depth of $1\sigma$ was then
produced and reconvolved to a circular beam of $66\farcs7\times66\farcs7$.
Because the central part of the original beam is well approximated by a
Gaussian, the fluxes measured do not require correction.

Channel maps of the cleaned reconvolved cube are shown in
Fig.~\ref{fig:channels}. Only the largest bubbles and shells are now visible,
due to the decreased spatial resolution compared to \markcite{PWBR92}PWBR92's
maps. The rotating disk pattern is still present, however, clearest at the
extreme velocity channels, and the warp is revealed by a change of the
kinematic major axis with radius. The most important feature of these maps is
the large extent of the \ion{H}{1} emission to the northwest of the galaxy.
The \ion{H}{1} extends to at least 16\arcmin\ from the center in certain
channels, almost four times $R_{25}$. The \ion{H}{1} cloud detected by
\markcite{PWBR92}PWBR92 is easily visible from $v_{\mbox{\scriptsize
    \sun}}=175$ to 195~\kms, while it was absent from individual channel maps
in the multi-configuration data cube.

\placefigure{fig:channels}

The clean data cube was used to produce moment maps with the AIPS task MOMNT.
These maps are shown in Fig.~\ref{fig:mom0}--\ref{fig:mom2}, superposed on an
optical image of HoII. The same cube, corrected for primary beam attenuation,
was used to calculate the global profile, summing the signal in each channel
using the moment~0 map as a mask. The global profile is shown in
Fig.~\ref{fig:global_d}.

The total \ion{H}{1} map shown in Fig.~\ref{fig:mom0}, being at
low-resolution, washes out a lot of the substructure visible in the maps of
\markcite{PWBR92}PWBR92, and provides us with an improved view of the
large-scale structure of the \ion{H}{1} in HoII. Undetected before, an
extended but low surface brightness component is clearly visible to the north
and northwest of the galaxy. The so-called \ion{H}{1} cloud detected by
\markcite{PWBR92}PWBR92 to the southwest of the galaxy is also visible and
appears to be part of the same structure as the \ion{H}{1} detected to the
northwest. Both components form a faint but large structure extending over the
entire northwest half of the galaxy. Given that, in addition, the \ion{H}{1}
appears compressed at the southeast edge of the galaxy, this gives rise to a
striking asymmetry between the southeast and northwest halves, the \ion{H}{1}
taking a comet-like appearance. This suggests that HoII may be affected by ram
pressure from an IGM. We will discuss this issue further in
\S~\ref{sec:ram+environment}.

\placefigure{fig:mom0}

The faint structure discovered to the northwest is very important since it
increases the area covered by \ion{H}{1} by almost a factor of 2 compared to
the work of \markcite{PWBR92}PWBR92. The total extent of the \ion{H}{1} now
reaches 16\arcmin\ or $4R_{25}$ on the northwest side of the galaxy, twice the
radius reached previously. This is a huge gain for studying the kinematics and
mass distribution of the system. 

The low-resolution velocity field shown in Fig.~\ref{fig:mom1} clearly shows a
differentially rotating disk pattern within at least the inner
$\approx7-8\arcmin$ in radius (albeit with an important warp). The situation
is less clear in the faint outer structure. There, while the velocities are
coherent with those slightly closer in (that is, there is no discontinuity
between the velocities in the faint structure and those within the
well-defined disk component), it is far from clear that we are dealing with a
stable structure, and there may well be large non-rotational motions. We
should thus expect difficulties to derive a rotation curve this far out. The
closed isovelocity contours at about 4\arcmin\ from the center do indicate,
however, that the rotation curve should peak around that radius.

\placefigure{fig:mom1}

The velocity dispersion field of HoII, shown in Fig.~\ref{fig:mom2}, displays
numerous quasi-circular regions where the dispersion peaks. These are
associated with expanding shells and were studied in details by
\markcite{PWBR92}PWBR92. The velocity dispersion in the disk of HoII is not
constant at 8--10~\kms, but increases steadily toward the center. Over most of
the optical disk, the dispersion is $\approx15$~\kms.  Since the rotation
velocity is $\sim25/\sin i$~\kms, where $i$ is the inclination of the disk,
the dynamical support provided by the dispersion is not negligible over a
large fraction of the disk, and it should be taken into account when modeling
the mass distribution (the ratio of pressure to rotational support goes
roughly as $\sigma^2/v^2$). This will be done in \S~\ref{sec:mass} by
correcting the derived rotation curve for asymmetric drift. The faint
structure to the northwest has much smaller velocity dispersions, typically
$\lesssim7$~\kms.

\placefigure{fig:mom2}

The global profile of the D-array data, shown in Fig.~\ref{fig:global_d}, is
very similar to those of the multi-configuration data shown in
Fig.~\ref{fig:global_multi}. The derived kinematical parameters are identical
($V_{\mbox{\scriptsize \sun}}=156\pm2$~\kms, $\Delta V_{50}=57\pm3$~\kms, and
$\Delta V_{20}=72\pm3$~\kms) and the total \ion{H}{1} flux is almost the same as
that of the corrected multi-configuration profile ($F_{\mbox{\scriptsize
    HI}}=267$~Jy~\kms). This flux translates into a total \ion{H}{1} mass of
$6.44\times10^8$~\msun. The agreement between the D-array data and the
corrected multi-configuration data supports the later and the correction
method proposed by J\"{o}rs\"{a}ter \& van Moorsel \markcite{jv95}(1995).
Indeed, the simple beam of the D-array data leads to a more reliable flux
determination. It also suggests that the D-array data is not missing much
flux. However, given the very extended nature of the \ion{H}{1} in HoII, this
can only be confirmed with reliable {\em low} resolution single-dish data.
Furthermore, the current extent of the \ion{H}{1} reaches the half-power beam
width (HPBW) of the VLA antennas ($\approx32\arcmin$) to the northwest of the
galaxy, so it is doubtful whether we could detect \ion{H}{1} farther out even
if it were there. Any follow-up observations of HoII with the VLA should use a
mosaicing strategy.

\placefigure{fig:global_d}
\section{Rotation Curve and Mass Modeling\label{sec:mass}}
\nopagebreak
\subsection{Determination of the Rotation Curve\label{sec:rotcur}}
\nopagebreak
The rotation curve was derived using the velocity field shown in
Fig.~\ref{fig:mom1} and the now standard algorithm ROTCUR in AIPS (see Begeman
\markcite{b87}1987). ROTCUR models a rotating disk as a series of concentric
inclined annuli, each annulus (radius $r$) having its own center
($x_0$,$y_0$), systemic velocity ($V_{sys}$), position angle (PA), inclination
($i$), and rotation velocity ($v_{rot}$). Given our beam of
$66\farcs7\times66\farcs7$, we have chosen annuli of width 60\arcsec\ starting
at $r=60\arcsec$\ (the inner 30\arcsec\ are thus unused). Data within
$30\arcdeg$ of the minor-axis were also rejected because of projection
effects. ROTCUR yields formal errors, but modeling the approaching and
receding sides of the galaxy separately often gives a more realistic idea of
the uncertainties.

As a first step, we want to fix the center and systemic velocity of the galaxy
to a common value for all the annuli. For this, we used only the inner
$7-8\arcmin$ of the velocity field, where the rotating pattern is clearest.
After various tests with all six parameters free and different initial
guesses, the systemic velocity was fixed to 156.8~\kms\ (heliocentric), in
agreement with the value obtained from the global profile, and the center was
fixed to ($\alpha=8\mbox{h} 13\mbox{m} 55\mbox{s}$, $\delta=+70\arcdeg
52\arcmin 52\arcsec$) (B1950), within one beam-width of the optical center.
Leaving these 3 parameters fixed, we then derived a rotation curve for the
inner parts of the velocity field only, using PA, $i$, and $v_{rot}$ to
minimize the residuals between the observed and modeled velocity fields. The
position angle then varies from 168 to $213\arcdeg$ and the inclination from
55 to $47\arcdeg$. This rotation curve is shown in Fig.~\ref{fig:rotcur_in}.
As the inclination we adopt is slightly higher than that of
\markcite{PWBR92}PWBR92, our rotation velocities are slightly lower
($\approx5$~\kms).

\placefigure{fig:rotcur_in}

As a final step, we derived the rotation curve using the entire velocity
field. Given that the constraints on a possible dark matter halo become
tighter with increasing radius, our aim was to extend the rotation curve to as
large a radius as possible. It is clear however that the rotation pattern is
very disturbed in the faint component to the northwest, making the derivation
of a rotation curve to large radii non-trivial. ROTCUR on its own did not
produce reliable fits in that region, so some manual intervention was required
to obtain good solutions. We guided ROTCUR to a fit yielding a reasonably
well-behaved rotation curve while keeping the residuals small.  Clearly, this
process is not entirely satisfactory, but it is the best that was possible
given the data in hand. The position angle is fairly well constrained, but it
is difficult to constrain reliably the inclination, particularly in the inner
parts, because of the degeneracy between $i$ and $v_{rot}$. Our adopted
rotation curve is shown in Fig.~\ref{fig:rotcur_out}, together with the
variations of $i$ and PA with radius. The values are tabulated in
Table~\ref{ta:rotcur}. As the data to the west are close to the minor-axis, it
is the data to the north in the faint outer structure which contribute most to
the rotation curve at large radii.

\placefigure{fig:rotcur_out}
\placetable{ta:rotcur}
\subsection{Asymmetric Drift Correction\label{sec:asym_drift}}
\nopagebreak
The rotation curve just derived shows a clear break at about $r=500\arcsec$,
where the faint outer structure begins. This casts doubts about the
reliability of the rotation curve past that point. However, as was pointed out
before, because of the high velocity dispersion in the disk of HoII, it is
likely that pressure support plays an important role. We must thus correct for
the asymmetric drift in order to obtain a more reliable estimate of the
circular velocity, truly representative of the underlying mass distribution.

The correction to the rotation curve can be written as
\begin{equation}
\label{eq:vc}
v_c^2=v_{rot}^2+\sigma_D^2, 
\end{equation}
where
\begin{eqnarray}
\label{eq:sigma_d}
\sigma_D^2 & = & -r\sigma^2\frac{\partial \ln(\rho\sigma^2)}{\partial r} \nonumber \\
           & = & -r\sigma^2\frac{\partial \ln(\Sigma\sigma^2)}{\partial r},
\end{eqnarray}
and $v_c$ is the circular velocity, $\sigma_D$ the asymmetric drift
correction, $\rho$ the volume density of \ion{H}{1}, and $\Sigma$ its surface
density. The last line is obtained by assuming an exponential distribution in
the vertical direction and a constant scaleheight.

To calculate the asymmetric drift correction, we thus need the surface density
and velocity dispersion radial profiles. These can be easily obtained from the
moment maps using azimuthal averages and the annuli adopted for the rotation
curve fit (we must also assume a constant position angle PA=168\arcdeg\ and
inclination $i=55\deg$). The resulting profiles are shown in
Fig.~\ref{fig:radial_profs}, where the surface densities have been
inclination-corrected. To obtain a smoother solution, we fit both $\sigma^2$
and $\Sigma\sigma^2$ with analytical functions. These profiles and the fitting
functions are also shown in Fig.~\ref{fig:radial_profs}.  $\sigma^2$ is well
fitted by a Gaussian $\sigma^2(r)=I_0\exp(-r^2/b^2)$, with
$I_0=242$~km$^2$~s$^{-2}$ and $b=392\arcsec$. $\Sigma\sigma^2$ is well
approximated by the function $\Sigma\sigma^2(r)=[I_0(R_0+1)]/[R_0+\exp(\alpha
r)]$, with $I_0=1350$~\msun~pc$^{-2}$~km$^2$~s$^{-2}$, $R_0=68\farcs3$, and
$\alpha=0.0159$~arcsec$^{-1}$. Using these functions, a smooth asymmetric
drift correction can be calculated easily.

\placefigure{fig:radial_profs}

The rotation curve of HoII corrected for asymmetric drift is shown in
Fig.~\ref{fig:circ_vel}, along with the actual correction applied and the
uncorrected rotation curve. The numerical values are tabulated in
Table~\ref{ta:rotcur}. The sharp increase of the rotation velocities seen at
$r\approx500\arcsec$ in the uncorrected curve is much weaker in the corrected
one; the jump is only 4--6~kms. This is because the asymmetric drift
correction is most important in the radial range 200--700\arcsec, where the
\ion{H}{1} surface density drops most rapidly. The corrected rotation curve,
our best estimate of the circular velocity curve, is not ideal for mass
modeling, but it should allow us to roughly constrain the mass distribution.

\placefigure{fig:circ_vel}
\subsection{Mass Modeling\label{sec:mass_model}}
\nopagebreak
The model we use to study the mass distribution of HoII is described in detail
in Carignan (\markcite{c85}1985) and Carignan \& Freeman
(\markcite{cf85}1985). This model calculates the mass contributions from the
luminous component (stars and \ion{H}{1}) and from a dark isothermal halo. To
model the mass contribution of the stellar disk, the $R$-band luminosity
profile (transformed into a $B$-band profile using the mean color
$(B-R)=0.767$; Swaters, private communication) is used with the assumption of
a constant, unknown $(M/L_B)_*$. An intrinsic flattening $q_0\equiv c/a=0.12$
is adopted for the stellar disk (Bottinelli et al.\ \markcite{bgpv83}1983).
For the \ion{H}{1} component, we use the radial surface density profile of
Fig.~\ref{fig:radial_profs}, multiplied by 4/3 to correct for the primordial
Helium content. The dark halo component is represented by an isothermal
sphere, described by two basic parameters: the core radius $r_c$ and the
one-dimensional velocity dispersion $\sigma$. The central density
$\rho_0=9\sigma^2/4\pi Gr^2_c$. The mass model therefore depends on three
parameters: the amplitude scaling of the disk $(M/L_B)_*$, the radial scaling
$r_c$, and the velocity scaling of the halo $\sigma$. In the fitting process,
each point is weighted according to its error.

The ``best-fit'' model is shown in Fig.~\ref{fig:mod_mass}. Two reference
radii need to be remembered: the optical radius, $\approx4$~kpc, and the
radius up to which velocity points are present on both sides of the galaxy,
$\approx10$~kpc. Up to the latter, we can consider the rotation curve to be
well defined. For $10\lesssim r\lesssim18$~kpc, the rotation curve is only
defined on the approaching side and one should be cautious when interpreting
it. This is especially true in light of the discussion in
\S~\ref{sec:ram+environment}. The parameters that best fit the observed
rotation curve are: the mass-to-light ratio of the stellar disk
$(M/L_B)_*=0.8$ and the core radius $r_c=10$~kpc and one-dimensional velocity
dispersion $\sigma=23$~\kms\ of the dark isothermal halo. These correspond to
a dark halo central density $\rho_0=0.00088$~\msun~pc$^{-3}$.

It is interesting to note that the model fits the rotation curve very well
both in the inner and outer parts, with the exception of the region around
10~kpc which marks the transition between the main body of the galaxy and the
\ion{H}{1} extension discovered in this study. If we restrain our analysis to
the region of the rotation curve which is well defined ($r\lesssim10$~kpc), it
can be noted that the mass distribution of HoII is quite different from what
is seen in most dwarfs and some late-type spirals, where dark matter dominates
even within the optical radius (Carignan \& Beaulieu \markcite{cb89}1989; 
C\^{o}t\'{e}, Carignan, \& Sancisi \markcite{ccs91}1991; C\^{o}t\'{e},
Carignan, \& Freeman \markcite{ccf00}2000). Here, dark matter
contributes only $\approx30$\% of the mass within 4~kpc.

Considering the whole extent of the rotation curve (out to $\approx18$~kpc),
we find a total mass of $6.3\times10^9$~\msun, of which about 80\% is dark
($M_{dark}/M_{lum}\approx4.5$). This yields a total dynamical
mass-to-light ratio
$(M/L)_{dyn}\approx16$. Now, similarly to what is seen in dwarfs and some
late-type spirals, there is more luminous mass in \ion{H}{1} than in stars
(about a factor of 2). Thus despite some differences between HoII and most
other dwarfs in the inner parts (luminous matter dominating within the optical
radius), the system as a whole is completely dark matter dominated at the last
point, {\em if we can trust the second half of the rotation curve}.

\placefigure{fig:mod_mass}
\section{Ram Pressure and The Environment of HoII\label{sec:ram+environment}}
\nopagebreak
\subsection{Ram Pressure\label{sec:ram}}
\nopagebreak
Ram pressure, whereby the gaseous medium of a galaxy is stripped by its
passage through a dense external medium, can be easily recognized by the
typical morphology it imposes on the disturbed ISM (see, e.g., Stevens,
Acreman, \& Ponman \markcite{sap99}1999; Mori \& Burkert \markcite{mb00}2000).
Although it is hard to prove without doubt that ram pressure is responsible
for any given disturbance in a galaxy, hot gas is known to be present in the
center of many clusters and groups (Mulchaey \markcite{m00}2000), so the ISM
of galaxies on radial orbits will be affected to varying degrees. Evidence for
ram pressure is also observed in many galaxies across a large range of
wavelengths (see White et al.\ \markcite{wffjs91}1991 for optical, infrared,
and X-ray observations of M86; Ryder et al.\ \markcite{rpda97}1997 for optical
and \ion{H}{1} observations of NGC7421; Gruendl et al.\ \markcite{gvdm93}1993
for optical and H$\alpha$ observations of NGC2276).

The threshold condition for ram pressure stripping can be written simply as 
\begin{equation}
\label{eq:ram}
\rho_{\mbox{\tiny IGM}}v^2\geq2\pi G\Sigma_{tot}\Sigma_{g}
\end{equation}
(Gunn \& Gott \markcite{gg72}1972), where the left side of the equation
represents the pressure exerted by an intragroup medium (IGM) of density
$\rho_{\mbox{\tiny IGM}}$ on a galaxy moving at a relative velocity $v$, and
the right side of the equation represents the restoring force of a disk with
total surface density $\Sigma_{tot}$ and gaseous (ISM) surface
density $\Sigma_{g}$.

The faint structure observed in the outer parts of HoII, to the northwest, is
suggestive of ram pressure. We can see from Fig.~\ref{fig:mom0} that the
\ion{H}{1} disk starts to be perturbed at a flux level of
$\approx4.7\times10^{19}$~atoms~cm$^{-2}$, corresponding to a deprojected
\ion{H}{1} surface density $\Sigma^i_{\mbox{\scriptsize
    HI}}=0.22$~\msun~pc$^{-2}$. In our azimuthally averaged surface density
profile (Fig.~\ref{fig:radial_profs}), this corresponds to a radius
$r\approx635\arcsec$. Assuming the optical disk to be exponential well past
the last measured point, we can extrapolate the optical surface brightness
profile of HoII to large radii. The fit to the outer parts of the disk by
Swaters \markcite{s99}(1999) yields a central surface brightness
$\mu_B(0)=22.28$~mag~arcsec$^{-2}$ and a scalelength $h_B=68\farcs1$, or
$\mu_B(r=635\arcsec)=32.4$~mag~arcsec$^{-2}$, corresponding to a deprojected
stellar surface density $\Sigma^i_{*}=0.0042$~\msun~pc$^{-2}$ for
$[(M/L)_B/(M/L)_B,{\mbox{\scriptsize \sun}}]=1$. This is negligible compared
to the \ion{H}{1} surface density.  We take $v\approx\sqrt{3}\sigma$ for the
velocity of HoII with respect to a putative IGM, where $\sigma$ is the
line-of-sight velocity dispersion of the M81 group of galaxy. Estimates of
$\sigma$ in the literature, based on the central members of the group, are
typically in the range $100-120$~\kms, so we adopt here the unweighted mean of
a few values, $\sigma\approx110$~\kms\ (see, e.g., Huchra \& Geller
\markcite{hg82}1982 and Garcia \markcite{g93}1993), leading to
$v\approx190$~\kms. Correcting the neutral hydrogen surface density for other
gaseous species, the IGM volume density required to strip the ISM of HoII at
$r\approx635\arcsec$ is, from Eq.~(\ref{eq:ram}), $n_{\mbox{\tiny
    IGM}}\gtrsim4.0\times10^{-6}$~atoms~cm$^{-3}$. We assume everywhere media
composed of 90\% hydrogen and 10\% He, yielding a mean mass per particle
$\mu=1.3m_{\mbox{\tiny p}}$ for neutral media and $\mu=0.75m_{\mbox{\tiny p}}$
for fully ionized media ($m_{\mbox{\rm p}}$ is the proton mass).

Eq.~(\ref{eq:ram}) does not take into account the presence of a dark
halo. However, if it interacts only through gravity and is not highly
flattened, the dark halo should not significantly affect ram pressure,
which acts mechanically and truly depends on the restoring force
inside the disk. In the case of spiral galaxies, Abadi, Moore, \&
Bower \markcite{amb99}(1999) show that the disk dominates the
restoring forces to very large radii. They find good agreement between
the analytic estimate of Gunn \& Gott \markcite{gg72}(1972) and their
simulations, except at small radii where their bulge dominates. The
dark halo will affect the evolution of the stripped material, in
particular its fall back, as the entire potential of the galaxy is
then important. Numerical simulations such as those discussed in
\S~\ref{sec:ram+shell_creation}, which are concerned with the entire
temporal evolution of the gas, do consider these effects (e.g.\
Quilis, Moore, \& Bower \markcite{qmb00}2000).

It is well known that the galaxies at the center of the M81 group are strongly
interacting, with \ion{H}{1} and optical bridges connecting the most massive
members, and tails of \ion{H}{1} extending to large radii (see, e.g., Cottrell
\markcite{c77}1977; Getov \& Georgiev \markcite{gg88}1988; Yun, Ho, \& Lo
\markcite{yhl94}1994). From the positions and radial velocities of the five
most prominent spiral members of the group, Huchra \& Geller
\markcite{hg82}(1982) derive a virial mass for the M81 group of
$1.13\times10^{12}$~\msun\ (corrected to our adopted distance). This of course
assumes that the group is bound, but since the virial crossing time of the
group is much less than a Hubble time ($\approx0.03\,H_0^{-1}$) and the
central members are strongly interacting, this should be verified. HoII lies
at the edge of the group, at a projected separation from M81 of about $8\fdg5$
on the sky, corresponding to 475~kpc at the distance of HoII. If one assumes
that 1\% of the virial mass of the group is dispersed uniformly in a hot IGM
within a sphere just enclosing HoII, the average IGM density within that
volume would be $1.4\times10^{-6}$~atoms~cm$^{-3}$. This is only three times
less than the IGM density required to strip the outer parts of the \ion{H}{1}
disk in HoII as calculated above (this assumes of course normal interacting
baryonic matter).

While we do not claim the above number to be realistic, it does
provide a benchmark with which to compare more sophisticated
calculations. Any interaction debris would be strongly concentrated
towards the center of the group and would be clumpy, reducing greatly
the IGM density where HoII lies (hot gas, of course, is smooth). We
have further considered a ``face-on'' encounter with the IGM, which is
probably not the case given the morphology observed. On the other
hand, since HoII is located at the very edge of the group, almost
detached from it, it is quite possible that HoII is infalling for the
first time, and it may well have a much larger velocity relative to
the group than that assumed here. In fact, the outer parts of the
group are probably not virialized. At a velocity of 190~\kms, HoII
would take $\approx2.4\times10^9$~yrs to reach the center of the M81
group. An infall scenario is further supported by the morphology of
the \ion{H}{1} gas, the tail pointing almost exactly away from the
center of the group. In fact, since the group velocity dispersion
adopted here is based only on a few large galaxies in the center of
the group, it is most likely a lower limit.  Including the fainter
members of groups, whose orbits are less affected by dynamical
friction and do not decay as rapidly, typically increases group
velocity dispersions by at least 50\% (up to a factor of 3; Zabludoff
\& Mulchaey \markcite{zm98}1998). This would reduce by at least a
factor of two the IGM density required for ram pressure
stripping. Most importantly, if one considers that HoII and/or other
galaxies outside of the group's core are bound to the M81 group, than
the virial mass assumed above is a serious underestimation, since only
the 5 innermost large members were used in the calculations. If one
assumes, for example, that the harmonic radius of the group
encompasses HoII, then the derived viral mass and IGM density (as well
as the crossing time) go up by a factor of at least 3.5. Any of these
factors can thus bring the required and derived IGM densities in
agreement with each other. In fact, from their spectroscopic survey of
poor groups, Zabludoff \& Mulchaey \markcite{zm98}(1998) find that
typical parameters of these groups are $R_{vir}\sim0.5\,h^{-1}$~Mpc,
$M_{vir}\sim0.5-1\times10^{14}\,h^{-1}$~\msun, and that only 10--20\%
of the mass is associated with individual galaxies. This leads to a
mean density within that radius of
$\approx3.7\times10^{-3}$~atoms~cm$^{-3}$ for the remaining matter
(for $H_0=75$~\kms~Mpc$^{-1}$). Of course, most of that is probably
non-baryonic, but on scales of the virial radius, the dominant
baryonic component of the mass in group is the IGM (Mulchaey
\markcite{m00}2000). And while this is strictly true for X-ray bright
groups only, where we can measure the gas content, it suggests at the
very least that ram pressure can not be rejected without further
analysis to explain the observed morphology of HoII at large radii.
\subsection{X-Ray Gas in Small Groups\label{sec:x-ray}}
\nopagebreak
Ponman et al.\ \markcite{pbeb96}(1996) have studied the X-ray properties of
Hickson's compact groups of galaxies (Hickson \markcite{h82}1982). A cursory
examination of the M81 group shows that it meets all three criteria set by
Hickson (indeed, it is not clear why the M81 group did not make it in the
list). Ponman et al.\ \markcite{pbeb96}(1996) detect X-ray emission in a large
number of groups, including spiral rich groups, and infer that hot extended
intragroup gas above their threshold of $L_x=10^{41.1}$~erg~s$^{-1}$ is
present in at least 75\% of the groups (excluding emission from individual
galaxies). The total X-ray luminosity of the groups does not correlate with
either the total number of galaxies or optical luminosity, but it does
correlate with the velocity dispersion and temperature (for Raymond \& Smith
1977 plasmas). For a velocity dispersion of $\sigma=110\pm10$~\kms, we obtain
from the correlations of the combined group and cluster data of Ponman
et al.\ \markcite{pbeb96}(1996) 
$L_x=10^{39.6\pm1.7}$~erg~s$^{-1}$ and $T_{\mbox{\tiny
    IGM}}=10^{-0.91\pm0.13}$~keV. However, there are indications that the
correlations for the groups and clusters may be significantly different, and
both of these numbers are probably underestimates. The large error on $L_x$ is
probably related to differences in the wind injection histories of the groups,
as this can have a significant effect in small potential wells. In any case,
it is clear that there could be a substantial amount of diffuse hot gas in the
M81 group.

Helsdon \& Ponman (\markcite{hp00a}2000a,\markcite{hp00b}b) carried out a
similar study on a large number of loose groups. Using MEKAL hot plasma fits
(Mewe, Lemen, \& van den Oord \markcite{mlo86}1986), they again obtained
correlations between the temperatures, velocity dispersions, and X-ray
luminosities of the groups. Using the loose groups correlations alone, we
obtain $L_x=10^{40.5\pm3.6}$~erg~s$^{-1}$ and $T_{\mbox{\tiny
    IGM}}=10^{-0.48\pm0.10}$~keV. Because the surface brightness profiles of
groups can be quite flat, most likely because of the effects of winds, the
reported luminosities are underestimates. For the coolest systems, the
luminosity within a virial radius can be underestimated by more than a factor
of two. 

Helsdon \& Ponman \markcite{hp00a}(2000a) also studied the surface brightness
profiles of the groups, essential if we are to estimate the typical amount of
hot gas present at the radius of HoII in a group like the M81 group.  They fit
the two-dimensional luminosity distributions with modified King functions
\begin{equation}
\label{eq:sb_prof}
S(r)=S_0(1+(r/r_c)^2)^{-3\beta+0.5},
\end{equation}
where $S_0$ is the central surface brightness, $r_c$ the core radius, and
$\beta$ measures the steepness of the profile. Most groups require two such
components to obtain a reasonable fit, but the three groups which are most
similar to the M81 groups are all well fitted by a single-component.
These are NGC~315, NGC~1587, and NGC~7777 (all three have four main members
and $\sigma=106-122$~\kms). Averaging their properties, we obtain
$L_x=10^{41.88\pm0.12}$~erg~s$^{-1}$, $T_{\mbox{\tiny IGM}}=0.80\pm0.16$~keV
(within the extracted radii), $r_c=6.6\pm5.8$~kpc, $\beta=0.73\pm0.56$, and
$a/b=1.16\pm0.21$ (elliptical fits). At large radii, Eq.~(\ref{eq:sb_prof})
becomes $S(r)=S_0(r/r_c)^{-6\beta+1}$, and in the isothermal
approximation the three-dimensional density profile then varies as
\begin{equation}
\label{eq:den_prof}
\rho(r)\propto(r/r_c)^{-3\beta}.
\end{equation}
At a distance of 475~kpc, it is clear that these groups have very
little IGM ($0.9\pm6.4\times10^{-4}$ of the density at the core
radius). However, if the $\beta$--$T_{\mbox{\tiny IGM}}$
trend observed by Helsdon \& Ponman \markcite{hp00a}(2000a) for the
higher quality (better statistics) two-component fits is
``universal'', then $\beta$ for these groups could be significantly
lower (by a factor of 2), leading to higher densities at large radii.

Another benchmark is provided by the study of poor groups of Zabludoff \&
Mulchaey \markcite{zm98}(1998). They report typical X-ray gas masses of about
$1\times10^{12}\,h^{-5/2}$~\msun\ for their groups, with virial radii
$\sim0.5\,h^{-1}$~Mpc. This leads to mean densities for the hot gas of
$\approx8.9\times10^{-5}$~atoms~cm$^{-3}$ within that radius. This is more
than enough to strip the ISM in the outer parts of a galaxy like HoII, but is
likely an overestimate for the density in the outer parts of the groups.
\subsection{Gas Survival in a Hot IGM\label{sec:survival}}
\nopagebreak
\subsubsection{The Effects of Shocks\label{sec:shocks}}
\nopagebreak
Given the low spatial velocity of HoII, the IGM may or may not be shocked.
With the range of temperatures allowed by the above correlations
($T_{\mbox{\tiny IGM}}=0.09-0.42$~keV, including the errors), the IGM sound
speed $c_{\mbox{\tiny IGM}}\approx(kT_{\mbox{\tiny IGM}}/\mu m_{\mbox{\tiny
    p}})^{1/2}$~\kms\ $\approx110-230$~\kms\ ($k$ is Boltzmann's constant;
$c_{\mbox{\tiny IGM}}\approx320$~\kms\ for $T_{\mbox{\tiny IGM}}=0.80$~keV).
Even if it is, the bow shock formed in front of HoII will barely affect
post-shock conditions. Rankine-Hugoniot jump conditions for adiabatic shocks
with $\gamma=5/3$ imply an increase (respectively decrease) in the post-shock
IGM density (respectively velocity) by a factor $M\approx v/c_{\mbox{\tiny
    IGM}}$. The effect is thus always less than a factor of two and will be
neglected in the following calculations. The estimates of viscous stripping
are in any case unaffected since the product $\rho_{\mbox{\tiny IGM}}v$ is
conserved. The IGM density required for ram pressure stripping (see
Eq.~\ref{eq:ram}) would be increased by up to 75\% for low IGM temperatures.

The \ion{H}{1} in HoII certainly will be shocked, with a shock velocity
$v_s\approx(n_{\mbox{\tiny IGM}}/n_c)^{1/2}\,v\approx0.4$~\kms\ at the
critical IGM density required for ram pressure stripping (McKee \& Cowie
\markcite{mc75}1975). However, this does not heat the gas significantly
($T\approx14\,v_s^2$~K).
\subsubsection{The Stripped Material\label{sec:stripped_gas}}
\nopagebreak
Any gas stripped from the main body of HoII must survive long enough in the
hot IGM to be observed. We follow here the discussion of Veilleux et al.\ 
\markcite{vbctm99}(1999) in the case of NGC~4388. Neglecting radiation and
magnetic fields, the evaporation timescale through thermal conduction for a
cloud embedded in the IGM is
\begin{equation}
\label{eq:evap_time}
t_{evap}\approx1\times10^3\, n_cR_{pc}^2\,(T_{\mbox{\tiny IGM}}/10^7 \mbox{\rm
  K})^{-5/2}\,(\ln \Lambda/30)\,\,\,\mbox{\rm yr}
\end{equation}
(Cowie \& McKee \markcite{cm77}1977), where $n_c$ is the mean hydrogen cloud
density in particles per cubic centimeter, $R_{pc}$ the cloud radius in
parsec, and $\Lambda$ the Coulomb logarithm. For a typical cloud
($n_c\approx1$~cm$^{-3}$, $R_{pc}\approx10$~pc), the evaporation timescale is
$6.2\times10^5-2.8\times10^7$~yrs (for $T_{\mbox{\tiny IGM}}=0.80$~keV,
$t_{evap}\approx1.2\times10^5$~yrs). The disturbed ISM observed in HoII
extends over about $7-8\arcmin$ in projection in the radial direction. At a
velocity of 190~\kms, it takes HoII about $3.6\times10^7$~yrs to cross that
distance. Given the strong dependence of the evaporation on the assumed
properties of the clouds and IGM, the timescales calculated seem compatible
with the observations, at least for the lower IGM temperatures.

One could also speculate that the (presumably) stripped gas originates from
the IGM itself. However, the cooling timescale for the IGM is
\begin{equation}
\label{eq:cool_time}
t_{cool}\approx4\times10^6\,(T_{\mbox{\tiny IGM}}/10^7 \mbox{\rm
  K})^{3/2}\,(n_{\mbox{\tiny IGM}}\, \mbox{\rm cm}^{-3})^{-1}\,\,\,\mbox{\rm yr}
\end{equation}
(Rangarajan et al.\ \markcite{rwef95}1995), which leads to
$t_{cool}\approx3.4\times10^{10}-3.4\times10^{11}$~yrs at the IGM density
required for ram pressure stripping ($t_{cool}\approx9.0\times10^{11}$~yrs for
$T_{\mbox{\tiny IGM}}=0.80$~keV). It is thus very unlikely that the gas
condensed out of the hot IGM.
\subsubsection{The HI Disk\label{sec:h1_disk}}
\nopagebreak
The gaseous disk of HoII as a whole will also be subjected to various
mass loss mechanisms
due to the presence of a hot IGM. We follow here the discussion of Nulsen
\markcite{n82}(1982) and consider in turn laminar viscous stripping, turbulent
viscous stripping, and thermal evaporation. These can remove material from the
disk even in regions not affected by ram pressure stripping.

Laminar viscous stripping leads to a mass loss
\begin{equation}
\label{eq:lam_loss}
\dot{M}_{visc}\approx(12/{\rm Re})\,\pi R_{HI}^2\rho_{\mbox{\tiny IGM}}v,
\end{equation}
where Re is the Reynolds number and $R_{HI}$ is the (maximum) radial extent of
the \ion{H}{1} disk (we use here $R_{HI}=16\arcmin=14.9$~kpc). The classical
treatment of viscosity requires that the ion mean free path in the IGM be
sufficiently small, $\lambda_{\mbox{\tiny IGM}}\lesssim R_{HI}$. In the
absence of magnetic fields, $\lambda_{\mbox{\tiny
    IGM}}\approx11\,(T_{\mbox{\tiny IGM}}/10^8\,\mbox{\rm
  K})^2\,(n_{\mbox{\tiny IGM}}/10^{-3}\,\mbox{\rm cm}^{-3})^{-1}$~kpc
$\approx0.3-6.4$~kpc at the critical density needed for ram pressure stripping
(Spitzer \markcite{s56}1956; $\lambda_{\mbox{\tiny IGM}}\approx23.7$~kpc for
$T_{\mbox{\tiny IGM}}=0.80$~keV). $\lambda_{\mbox{\tiny IGM}}\lesssim R_{HI}$
is thus only marginally satisfied, although tangled magnetic fields would
reduce it more. Laminar flows require $\mbox{\rm Re}\lesssim30$, but
$\mbox{\rm Re}=2.8\,(R_{HI}/\lambda_{\mbox{\tiny IGM}})\,(v/c_{\mbox{\tiny
    IGM}})\approx5-240$ (Batchelor \markcite{b67}1967; $\mbox{\rm Re}=1$ for
$T_{\mbox{\tiny IGM}}=0.80$~keV). The flow will thus unlikely be laminar, but
turbulent, except possibly for a small range of intermediate
$T_{\mbox{\tiny IGM}}$.

In a turbulent flow, Kelvin-Helmholtz instabilities will also generate mass
loss at a rate
\begin{equation}
\label{eq:turb_loss}
\dot{M}_{turb}\approx\pi R_{HI}^2\rho_{\mbox{\tiny IGM}}v
\end{equation}
(Nulsen \markcite{n82}1982). At the critical density required for ram pressure
stripping, $\dot{M}_{turb}\approx1.0\times10^{-2}$~\msun~yr$^{-1}$. It would
thus take $6.4\times10^{10}$~yrs, much more than a Hubble time, to deplete the
entire \ion{H}{1} content of HoII with turbulent viscous stripping only. This
process is thus unlikely to compete with ram pressure stripping, which removes
the gas bodily from the galaxy. We recall that the time required for HoII to
reach the center of the M81 group at a velocity of 190~\kms\ is 25 times
smaller, $t_{cross}\approx2.4\times10^9$~yrs. This suggests that if ram
pressure was inefficient, HoII could cross the group many times without
loosing its ISM.

The mass loss due to evaporation through thermal conduction is
\begin{equation}
\label{eq:evap_loss}
\dot{M}_{evap}\approx4\pi R_{HI}^2\rho_{\mbox{\tiny IGM}}c_{\mbox{\tiny IGM}}\phi_sF(\sigma_0)
\end{equation}
(Cowie \& McKee \markcite{cm77}1977), where $\phi_s\approx1$ and
$\sigma_0=1.84\,\lambda_{\mbox{\tiny IGM}}/R_{HI}\phi_s\approx0.04-0.80$
($\sigma_0=2.93$ for $T_{\mbox{\tiny IGM}}=0.80$~keV).
$F(\sigma_0)\approx2\sigma_0$ for $\sigma_0\lesssim1$, leading to
$\dot{M}_{evap}\approx1.7\times10^{-3}-7.8\times10^{-2}$~\msun~yr$^{-1}$
($4.0\times10^{-1}$~\msun~yr$^{-1}$ for $T_{\mbox{\tiny IGM}}=0.80$~keV) and a
depletion timescale for the \ion{H}{1} in HoII of
$8.3\times10^9-3.7\times10^{11}$~yrs ($1.6\times10^9$~yrs for $T_{\mbox{\tiny
    IGM}}=0.80$~keV). Except for very high $T_{\mbox{\tiny IGM}}$, thermal
evaporation is thus about as efficient as turbulent stripping to remove gas
from the disk of HoII. However, even combined, these two mechanisms would
still require a large fraction of a Hubble time to deplete the \ion{H}{1}
entirely.

We note that the ``formalism'' of turbulent viscous stripping applied here to
the disk of HoII can equally be applied to our typical \ion{H}{1} cloud,
considered in the previous subsection. The timescale for complete viscous
stripping can then be written as
$t_{turb}=M_c/\dot{M}_{turb}\approx4/3\,(R_c/v)(n_c/n_{\mbox{\tiny IGM}})$,
where $M_c$ and $R_c$ are the total mass and radius of the cloud,
respectively. At the IGM density required for ram pressure stripping, this
leads to $t_{turb}\approx1.7\times10^{10}$~yrs, while we derived
$t_{evap}\approx6.2\times10^5-2.8\times10^7$~yrs
($t_{evap}\approx1.2\times10^5$~yrs for $T_{\mbox{\tiny IGM}}=0.80$~keV).  For
the stripped material, therefore, evaporation is much more efficient than
viscous stripping at removing gas.

Given the absence of extended X-ray emission associated with HoII
(Zezas, Georgantopoulos, \& Ward \markcite{zgw99}1999), it is unlikely
to possess a gaseous halo protecting it from the IGM, as suggested by
Veilleux et al.\ \markcite{vbctm99}(1999) in the case of
NGC~4388. However, if the HoII, UGC~4483, and Kar~52 triplet or the
NGC~2403 subgroup (see \S~\ref{sec:m81_group} below) were to possess
such a halo, then it could possibly offer some protection.
\subsection{M81 and NGC~2403 Groups\label{sec:m81_group}}
\nopagebreak
While the morphology of the \ion{H}{1} in HoII (Fig.~\ref{fig:mom0}) is
reminiscent of that expected from ram pressure stripping, it could perhaps
also be caused by an interaction. As mentioned before, the main galaxies at
the center of the M81 group are known to be strongly interacting (see, e.g.,
Yun et al.\ \markcite{yhl94}1994), and the ``cometary'' tail of \ion{H}{1}
could then represent material tidally stripped from the main body of HoII.
However, HoII lies about 475~kpc from M81 (in projection), so it would take
it at least one fifth of a Hubble time to reach the center of the group at a
velocity of 190~\kms.

Figure~1 of van Driel et al.\ \markcite{dkbh98}(1998) shows a map of the
galaxies in the M81 group (their scale bar would be 0.4~Mpc at our adopted
distance to HoII). HoII, along with Kar~52 (better known as M81 Dwarf~A) and
UGC~4483, appears to be part of a subsystem of three dwarf irregular galaxies
to the northwest of the group's core. Kar~52 lies about 30~kpc to the
northeast ($\Delta V_r\approx40$~\kms), twice the maximum radial extent of the
\ion{H}{1}, and UGC~4483 is about 105~kpc to the southeast ($\Delta
V_r\approx0$~\kms). If HoII is interacting, it must be with one of these two
galaxies. Figure~7 in Karachentsev et al.\ (\markcite{ketal00}2000) shows a
larger map including the dwarf irregular galaxy NGC~2366, the dwarf spheroidal
DDO~44, and the large spiral NGC~2403, to the west and southwest of HoII.
These are sometime associated with the M81 group (e.g.\ de Vaucouleurs
\markcite{d75}1975) or may form a subgroup associated with NGC~2403.

Bremnes, Binggeli, \& Prugniel \markcite{bbp98}(1998) present optical CCD
images of both Kar~52 and UGC~4483. They are much smaller than HoII
($\approx15-20$\% at $R_{25}$) and are at least 4--5 magnitudes fainter. Both
galaxies also have rather irregular optical morphologies, but neither shows
obvious signs of interaction. UGC~4483 was recently studied with HST by
Dolphin et al.\ (\markcite{detal01}2001). Modeling of the color-magnitude
diagram argues for a roughly constant star formation rate (SFR) of
$1.3\pm0.2\times10^{-3}$~\msun~yr$^{-1}$, except for a young star cluster to
the north ($\lesssim100$~Myr), and the tip of the red giant branch yields a
distance of $3.2\pm0.2$~Mpc. Kar~52 shows no sign of star formation (Miller \&
Hodge \markcite{mh94}1994), but no resolved stellar population study is
available.

Kar~52 was first detected in \ion{H}{1} by Lo \& Sargent (\markcite{ls79}1979)
and was further studied by Sargent, Sancisi, \& Lo (\markcite{ssl83}1983). The
\ion{H}{1} is distributed in a lumpy ring, incomplete to the northwest, which
appears supported by turbulence (very little rotation is observed). If not for
the fact that the broken ring appears to surround the optical galaxy, the
morphology would again be very reminiscent of ram pressure. No \ion{H}{1} is
detected in the direction of HoII (to a column density of
$\approx10^{20}$~cm$^{-2}$), and there is no evidence of recent star
formation. VLA and Westerbork \ion{H}{1} observations of UGC~4483 are
available in van Zee, Skillman, \& Salzer \markcite{zss98}(1998) and
the WHISP database, respectively. The \ion{H}{1} is centrally
concentrated and the inner parts roughly follow the optical
morphology, but a more extended, faint, and patchy component
elongated NW--SE is also present. Only mild evidence of compression is
observed at low levels on the southeast side, both in the total
\ion{H}{1} map and the velocity field.

Our adopted distance to HoII is identical to that of UGC~4483 (Dolphin et al.\ 
\markcite{detal01}2001), but also to the distances of NGC~2403 (Freedman \&
Madore \markcite{fm88}1988) and DDO~44 (Karachentsev et al.\ 
\markcite{ketal99}1999). This suggests that HoII really belongs to the
NGC~2403 subgroup, together with Kar~52, UGC~4483, NGC~2366, and DDO~44.
Karachentsev et al.\ (\markcite{ketal00}2000) argue along the same lines. They
show that the distance to the core of the M81 group is about 0.5~Mpc larger
than that to the NGC~2403 subgroup, and that the mean radial velocity of the
M81 group is {\em smaller} than that to the NGC~2403 subgroup. They thus
suggest that the M81 and NGC~2403 groups (including HoII) are moving towards
each other and estimate a relative velocity of 110--160~\kms.

The lack of evidence for interaction in the triplet of galaxies
including HoII do not support 
this as the likely origin for its \ion{H}{1} morphology. In fact, the
\ion{H}{1} observations of both Kar~52 and UGC~4483 are somewhat reminiscent
of ram pressure. More sensitive \ion{H}{1} observations of the entire region
around HoII, Kar~52, and UGC~4483 are necessary to clarify this issue.
Distances to individual galaxies in the M81 group and NGC~2403 subgroup argue
that HoII is part of the latter and may be moving towards M81 at a velocity of
110--160\kms, similar to or somewhat smaller than our adopted velocity of
190~\kms. Our calculations concerning ram pressure stripping thus do not
depend strongly on whether HoII is bound to the M81 group or belongs to the
NGC2403 subgroup and is infalling towards M81. In the latter case, an IGM more
dense than what we have assumed (by up to a factor of 3) would be required to
strip the ISM of HoII through ram pressure.

We note that \ion{H}{1} synthesis observations of many other galaxies in the
M81 group are available in the literature (see, e.g., Westpfahl et al.\ 
\markcite{wcat99}1999). HoI and Kar~73, located closer to the group's core
(but still well outside of the M81, M82, and NGC~3077 triplet), display
disturbed morphologies not unlike those expected from ram pressure. At least
for HoI, it appears as if the galaxy is falling towards the center of the
group. Ott et al.\ (\markcite{owbddk01}2001) discuss the possibility of ram
pressure in this galaxy succinctly. Both HoI and Kar~73 were observed with the
B, C, and D arrays of the VLA, but reanalyzing the D-array data only would be
worthwhile, as we have done for HoII.
\section{The Creation of Shells and Supershells\label{sec:shells}}
\nopagebreak
\subsection{SNe and Stellar Winds\label{sec:sne}}
\nopagebreak
\markcite{PWBR92}PWBR92 studied in details the numerous bubbles and shells
present in the disk of HoII. Their main goal was to test whether these could
be formed by SNe and stellar winds. We will briefly review their and following
works here and highlight weaknesses of this scenario in the particular case of
HoII.

\markcite{PWBR92}PWBR92 identified and cataloged a large number of holes in
HoII, measuring their positions, apparent sizes, ellipticities, and expansion
velocities. These then yield simple estimates of the kinematic ages,
previously enclosed \ion{H}{1} masses, and creation energies of the holes.
\markcite{PWBR92}PWBR92 present various correlations between these properties,
arguing for SNe and stellar winds as progenitors for the holes.  However, many
correlations presented are in fact a reflection of the physical properties of
the host galaxy rather than of the creation mechanism of the holes, and many
correlations that do depend on the creation mechanism could equally arise from
a number of processes where the events are spread in time, cluster around a
given energy, and where the most extreme events are rare. Support for the SNe
and stellar wind scenario comes mainly, in our opinion, from the fact that at
least some holes appear to be (roughly) spherically expanding and/or have a
complete shell. Both of these properties point to internal pressure-driven
events. Unfortunately, it is hard to gauge what fraction of the holes analyzed
by \markcite{PWBR92}PWBR92 really show such convincing evidence.

Contrary to what is claimed by \markcite{PWBR92}PWBR92, we do no believe that
their H$\alpha$ image (their Fig.~21) strongly supports the SNe and stellar
wind scenario. The H$\alpha$ emission is not preferentially located at the
edges of large holes, many walls in between holes showing no H$\alpha$
emission at all, neither does it preferentially fill small holes, many of them
showing no H$\alpha$ either. Those properties are only verified in the very
center of HoII. Furthermore, if SNe are indeed responsible for the formation
of the largest holes, then these should by filled by hot X-ray gas
($\sim10^6$~K), which has a long cooling time. However, pointed $ROSAT$ PSPC
and HRI observations of HoII analyzed by Zezas et al.\ \markcite{zgw99}(1999)
reveal only a single unresolved X-ray source, coincident with a large
\ion{H}{2} region, and variable over a wide range of timescales. The bubbles
in HoII therefore appear to be devoid of hot gas, and the X-ray emission which
is seen probably arises from a compact accreting object. Kerp \& Walter
(\markcite{kw01}2001) reanalyzed the $ROSAT$ data, striving to reach the
faintest fluxes possible. The X-ray emission observed is not preferentially
associated with \ion{H}{2} regions or \ion{H}{1} holes, and a mixed bag of
objects is detected at various locations within HoII. Of the 13 sources
studied, 5 are located within \ion{H}{1} holes, 4 in high column density
regions, and 4 outside the stellar body of the galaxy. The latter detections
are perhaps the most significant since they are likely SN remnants (SNRs) or
X-ray binaries and suggest that star formation (SF) took place in the past
well outside the regions where it is occurring now.

The issue of energy injection in the ISM is also crucial to the SNe and
stellar wind scenario. Tongue \& Westpfahl \markcite{tw95}(1995) present
multi-frequency radio continuum observations of HoII, allowing them to
identify the dominant mechanism of radio emission in each structure observed.
Of eight unresolved sources identified, only 3 have non-thermal emission
probably associated with SNRs. Some diffuse emission consistent with
non-thermal disk emission is also observed. The three probable SNRs are very
strong, however, and they dominate the total flux. Tongue \& Westpfahl
\markcite{tw95}(1995) show that the total SN rate required to form the bubbles
in HoII (taken as the sum of the energy required to form the bubbles divided
by the largest kinematic age and the typical energy output of a SNe) is in
agreement with the total SN rate derived from the radio continuum emission
(using three different methods, each associated with one emission mechanism).
This suggest that the energy input from SNe alone can account for the
substructure observed in the \ion{H}{1} disk. However, as correctly pointed
out by Tongue \& Westpfahl \markcite{tw95}(1995), the radio continuum maps
correlate very well with the H$\alpha$ image of \markcite{PWBR92}PWBR92.  This
in turn means that they do {\em not} correlate well with the \ion{H}{1}
distribution. Indeed, the radio continuum emission is confined to the very
inner parts of the \ion{H}{1} distribution, away from most of the bubbles and
shells. This suggest that the large energy output from SNe is not deposited at
the right locations in the disk of HoII, hardly helping to identify the
formation mechanism of the holes. Using numerical simulations, Mashchenko \&
Silich \markcite{ms95}(1995) also tried to model the evolution of shells
expanding under a local energy source in HoII, but numerous inconsistencies
with the observations remained, in particular with respect to the elongation
of the shells.

If the SNe scenario is right in the case of HoII, then about 25\% of the holes
cataloged by \markcite{PWBR92}PWBR92 require more than 10 SNe to be created
(up to 200).  Given the ages of the bubbles ($10^7-10^8$~years), a significant
population of A, F, and perhaps B-type stars should be left at the center of
these holes (assuming a normal initial mass function). Rhode et al.\ 
\markcite{rswr99}(1999) obtained deep broadband $BVR$ images of HoII to look
for the remaining stellar clusters in all of the bubbles requiring at least
one SNe. Unfortunately, useful constraints could only be derived for
$\lesssim30\%$ of the holes, as the background light in the center of the
galaxy is high and some holes need only harbor a small central cluster.  In
most cases where useful constraints are derived, the expected stellar clusters
are simply not there, putting strong doubts on the validity of the SNe and
stellar wind scenario for the formation of (at least) these holes. Stewart et
al.\ \markcite{setal00}(2000) also looked for the presumed SF and stellar
populations at the origin of the holes, this time using far-ultraviolet (FUV)
imaging (as well as $UBR$ and H$\alpha$). FUV emission is sensitive to massive
SF over a timescale comparable to the kinematic ages of the \ion{H}{1} holes
($\lesssim100$~Myr), thus allowing to directly search for the holes'
progenitors (H$\alpha$ is only sensitive to very recent SF, $\lesssim5$~Myr).
Stewart et al.\ \markcite{setal00}(2000) find that the FUV emission is
concentrated along the high density edges of the \ion{H}{1} holes; no FUV
knots are observed at the center of the shells. Only 3 of the 51 holes
identified by \markcite{PWBR92}PWBR92 appear to have associated FUV emission.
In those three cases, the lower limit to the estimated mechanical energy
output from the associated stellar clusters is systematically higher than that
necessary to create the holes (but consistent in 2 cases).  Furthermore, very
little FUV emission is detected outside the inner $2-3\arcmin$, clearly
indicating that the stellar wind and SNe scenario is not viable outside that
region.

While the aforementioned studies seem to suggest that SF does depend on local
conditions (i.e.\ the amount of shear), and while they weakly support some
kind of sequential SF, they provide no smoking gun evidence that stellar winds
and SNe are at the origin of the \ion{H}{1} holes observed in HoII. The
problem is most acute for the most energetic holes, which are generally
located well outside of the optical extent of the galaxy. To a certain extent,
the results of Tongue \& Westpfahl \markcite{tw95}(1995), Rhode et al.\ 
\markcite{rswr99}(1999), and Stewart et al.\ \markcite{setal00}(2000) were to
be expected since a cursory examination of the \ion{H}{1}, optical broadband,
and H$\alpha$ images of HoII shows that many holes are located in low surface
brightness regions of the disk, where no SF is expected or appears to be
taking place.

We note that despite the fact that we have revised \markcite{PWBR92}PWBR92's
fluxes down by 23\%, the problem of the high creation energies of the
\ion{H}{1} holes remains entirely. The estimated creation energy of a hole
depends on the surface density at that point only quasi-linearly ($E_c\propto
n_{\mbox{\tiny HI}}^{1.12}\,$; Heiles \markcite{h79}1979), therefore decreasing
the values published by \markcite{PWBR92}PWBR92 by 25\% at most.
\subsection{Alternative Explanations\label{sec:alternatives}}
\nopagebreak
While we do not wish to challenge the general relevance of SNe and stellar
winds in shaping the ISM, the numerous shortcomings of this scenario in the
specific case of HoII suggest that additional energy inputs and/or mechanisms
are required to explain the observed substructure of the \ion{H}{1},
particularly regarding the largest and most energetic holes located in the
outer parts of the disk. Such a mechanism, ram pressure, is suggested by our
reanalysis of \markcite{PWBR92}PWBR92's data. Of course, numerous other
processes exist which can explain the formation of bubbles and shells in
galaxies, or can help to reconcile theoretical models with the observations
just described. Rhode et al.\ \markcite{rswr99}(1999) discuss these at some
length, so we only list them briefly here, and then focus in
\S~\ref{sec:ram+shell_creation} on how ram pressure may provide yet another
explanation in the particular case of HoII.

The energies necessary to create the holes were calculated using the
spherically-symmetric SNR model of Chevalier \markcite{c74}(1974), which
assumes a uniform ISM. More complex models for both the SN evolution and/or
the ISM could lead to lowered energies and decrease the flux expected from the
progenitor clusters. A (extremely) top-heavy initial mass function (IMF) could
also produce the required amount of SNe without an associated population of
lower mass stars. Similarly, the blast waves associated with gamma-ray bursts,
if caused by the collapse of a single massive star to a black hole, could
create large holes in the ISM without requiring large stellar clusters.  All
these mechanism, however, still require massive SF to occur wherever there are
holes, and this seems unlikely in the case of HoII. Other mechanisms bypass
this requirement.

If the \ion{H}{1} in galaxies is fractal (see, e.g., Westpfahl et al.\ 
\markcite{wcat99}1999 for a study of the \ion{H}{1} structure in M81 group
galaxies), then holes may occur naturally through the same processes which
create the fractal structure. A fractal \ion{H}{1} would lead to significant
changes in the energies required to create holes through SNe, as it would
affect all heating and cooling timescales discussed in the previous sections.
Elmegreen \& Hunter \markcite{eh00}(2000) also suggest that if giant
\ion{H}{2} regions observed in galaxies such as HoII are truly overpressured,
then they should represent only a small fraction of the true population of
\ion{H}{2} regions ($\approx7\%$; the timescale to reach pressure equilibrium
being more than an order of magnitude larger than their present age). The
unobserved old and faint \ion{H}{2} regions could give rise, by merging, to
large shells filling most of the disks and without recognisable associated
clusters. Finally, the \ion{H}{1} holes in the disk of HoII could be caused by
ionization from an external source, such as the metagalactic UV radiation
field (Bland-Hawthorn \markcite{b99}1999), especially if the large scaleheight
argued for by \markcite{PWBR92}PWBR92 is confirmed.

High-velocity clouds (HVCs) also have the power to create large (cylindrical)
holes in the ISM when impacting a disk. Holes in our galaxy and others have in
fact been convincingly linked with high-velocity gas (see, e.g., Heiles
\markcite{h84}1984 for the Galaxy; van der Hulst \& Sancisi
\markcite{hs88}1988 for M101). Tenorio-Tagle (\markcite{t80}1980,
\markcite{t81}1981) discusses their formation and expansion. Rhode et al.\ 
\markcite{rswr99}(1999) explored this possibility in the case of HoII. They
reexamined the low-resolution datacube of \markcite{PWBR92}PWBR92 and
identified one possible HVC. Although the detection is not statistically
significant, the properties of the candidate cloud are interesting. Its
emission extends over three velocity channels (the minimal criterion set by
them) and about 90\arcsec\ spatially, with a total flux of 4.9~Jy~\kms,
yielding a total \ion{H}{1} mass of $1.2\times10^7$~\msun. The cloud is
located 17\arcmin\ to the southeast of HoII, at the edge of the VLA primary
beam, and in the opposite direction from where we detect extended emission. It
has a velocity relative to HoII of $\approx65$~\kms, giving it a kinetic
energy of $\approx5\times10^{53}$~ergs in HoII's rest frame. This is more than
enough to create any of the holes observed. Clearly, many more less massive
clouds, with the possibility of releasing $\lesssim10^{52}$~ergs in the disk
of HoII, could remain undetected.
\subsection{Ram Pressure and the Creation of Shells\label{sec:ram+shell_creation}}
\nopagebreak
Ram pressure is a new ingredient that must be considered when studying the
large- and small-scale structure of HoII. Ram pressure can create holes in a
dense gaseous disk where local minima in the surface density exist. Together
with viscous stripping, it also provides an efficient mechanism to enlarge
pre-existing holes, created by SNe or otherwise, by ablating their edges. In
the case of HoII, it could thus explain the surprisingly large creation
energies of the holes and the lack of observational signatures expected from
SNe and stellar wind scenarios. Unfortunately, no simulations or detailed
calculations are available to verify the validity of this process
quantitatively. The formation of holes through ram pressure should thus be
modeled carefully before making any further claims.

It is interesting to note that in clusters with a deep potential well and/or a
high gas density, the SFR of infalling disk galaxies can increase by up to a
factor of two early on due to ram pressure compression of the
molecular gas (Fujita \& 
Nagashima \markcite{fn99}1999). The SFR decreases rapidly as the galaxy moves
closer to the cluster center, however, due to stripping. As expected, the
effect is less significant for shallower potential wells and/or lower gas
densities. The tidal field of a cluster as a whole is also very efficient at
triggering SF in disk galaxies, up to many core radii (Byrd \& Valtonen
\markcite{bv90}1990). It is unclear however how efficient these mechanisms can
be in groups, where despite a less extreme environment the least massive disk
galaxies (such as dwarf irregulars) could still be affected. Mori \& Burkert
(\markcite{mb00}2000) do consider the effects of ram pressure and turbulent
viscous stripping on dwarfs in clusters, but they consider a pressure
supported extended hot ISM, so their results are not directly applicable to
our situation. According to Mori \& Burkert (\markcite{mb00}2000), if HoII
were to have a hot gaseous halo, it would not be stripped.

It should be possible observationally to distinguish ram pressure from
internal, pressure-driven events such as SNe and stellar winds. One may expect
the holes created from ram pressure to have a structure somewhat similar to
that of a ``bullet-hole'', like that caused by the impact of a HVC. A
one-sided shell should be created first, soon followed by an expanding
cylindrical hole (e.g.\ Tenorio-Tagle \markcite{t80}1980). In the case of ram
pressure, however, one would also expect the characteristic tails of warm
\ion{H}{1} in pressure equilibrium with the intracluster medium (ICM) or IGM.

Quilis et al.\ (\markcite{qmb00}2000) present three-dimensional,
self-consistent, $N$-body and hydrodynamical simulations of disk galaxies
(with a bulge and dark matter halo) moving through a dense ICM. Not only do
they consider different relative velocities, infall geometries, and ICM
densities, but also different ISM distributions in the galaxy. Of particular
interest to us, they consider gaseous disks with a large central hole or with
numerous small holes distributed ramdomly throughout the inner parts. These
mimic the central depletion of \ion{H}{1} observed in many spiral galaxies and
the numerous bubbles, shells, and holes present in spirals and dwarfs, be they
created by SNe and stellar winds or otherwise. Compared to simulations with a
homogeneous ISM, ram pressure stripping and viscous stripping are then much
more efficient, removing material from the edges of the holes and preventing
material from falling back on the galaxy. The morphologies obtained in the
simulations compare advantageously with the \ion{H}{1} distribution of HoII.
We note that hydrodynamic codes unable to model viscosity and turbulence yield
much smaller mass losses (e.g.\ Abadi et al.\ \markcite{amb99}1999).
In all simulations, however, the timescale for gas depletion is much smaller
than the crossing timescale of the cluster, so ram pressure acts very rapidly.
This is consistent with HoII being on the outskirts of the M81 group, probably
infalling for the first time.

The conditions in the simulations of Quilis et al.\ (\markcite{qmb00}2000) are
not directly comparable to those in the M81 group or poor groups in general.
However, while galactic velocities and intergalactic medium densities are both
lower in groups (by an order of magnitude), typical disk surface densities in
dwarfs are also much lower than in giant spiral galaxies. The inequality in
Eq.~\ref{eq:ram} may thus still be verified for dwarfs. The question is
whether or not stripping occurs at a sufficiently small radius (or high
\ion{H}{1} surface density) to be detectable with current instruments.
Simulations along these lines, taking into account the highly inhomogenous
nature of the ISM in dwarfs, would be welcome.
\section{Conclusions\label{sec:conclusions}}
\nopagebreak
With low resolution but sensitive VLA observations, we have shown that the
large-scale structure of the \ion{H}{1} disk in HoII is comet-like. Since HoII
does not appear to be interacting with its neighbors and may be infalling
towards to core of the M81 group, it is probably undergoing ram pressure
stripping from an IGM. About 1\% of the virial mass of the group must be
contained in an IGM if the outer parts of HoII are to be stripped, consistent
with X-ray observations of small groups. The length and morphology of the
\ion{H}{1} tail observed point to the additional effects of turbulent viscous
stripping and thermal evaporation.

Ram pressure is interesting with regard to the creation of the \ion{H}{1}
holes in HoII. Since the expected traces of the implied SF are not observed,
the creation energies of the holes through SNe and stellar winds must be
systematically overestimated (or one must reject altogether this formation
scenario). Ram pressure offers a mechanism to enlarge pre-existing holes, no
matter how they were created, and thus lower their creation energies. A proof
of the existence of a sufficiently dense IGM remains, however, the missing
element to support this suggestion.

While detecting a diffuse hot IGM of density
$10^{-6}-10^{-5}$~atoms~cm$^{-3}$ would be very hard, especially in
the M81 group given its 
large extent on the sky, it may be easier to look for other signatures of its
presence. These include leading bow shocks (for galaxies moving supersonically
through the IGM), gravitationally focused wakes, and the ram pressure stripped
material itself (see Stevens et al.\ \markcite{sap99}1999). These may be
detectable with the new generation of X-ray telescopes ({\em XMM} and {\em
  Chandra}) and could also be used to constrain the orbital structure
(anisotropy) of the group (Merrifield \markcite{m98}1998). HoII is a prime
target for a search since it lies on the outskirt of a poor group with a
presumably low IGM temperature. Contrary to expectations, the effects of ram
pressure stripping are most easily visible in low surface brightness, cool,
poor groups and clusters. In these, galaxies are able to retain a substantial
fraction of their ISM, leading to prominent tails (see, e.g., Model~1b of
Stevens et al.\ \markcite{sap99}1999)

Deep \ion{H}{1} observations of the region around HoII, Kar~52, and UGC~4483
are also crucial. They could reveal evidence of interactions that were missed
by the present observations and which could also explain HoII's large-scale
morphology. They could also reveal or rule out the presence of dense, cold
material in the vicinity of HoII (e.g.\ HVCs), which may be related to
the creation of some of the holes. Observations along these lines are
underway with the D configuration at the VLA.
\acknowledgments
We thank D.\ Puche for making available the reduced data of HoII. We also
thank K.\ C.\ Freeman and J.\ H.\ van Gorkom for useful discussions. M.\ B.\ 
acknowledges the support of a Canadian NSERC Undergraduate Student Research
Award, an Australian DEETYA Overseas Postgraduate Research Scholarship, and a
NSERC Postgraduate Scholarship during various part of this
work. Support for this work was also provided by NASA through Hubble
Fellowship grant HST-HF-01136.01 awarded by Space Telescope Science
Institute, which
is operated by the Association of Universities for Research in
Astronomy, Inc., for NASA, under contract NAS~5-26555. The Digitized
Sky Surveys were produced at the Space Telescope Science Institute under U.S.
Government grant NAG W-2166. The images of these surveys are based on
photographic data obtained using the Oschin Schmidt Telescope on Palomar
Mountain and the UK Schmidt Telescope. The plates were processed into the
present compressed digital form with the permission of these institutions.
The NASA/IPAC Extragalactic Database (NED) is operated by the Jet Propulsion
Laboratory, California Institute of Technology, under contract with the
National Aeronautics and Space Administration. This project also made use of
the LEDA (www-obs.univ-lyon1.fr) and WHISP (http://www.astro.rug.nl/~whisp/)
databases.

\clearpage
%
%
%
\figcaption[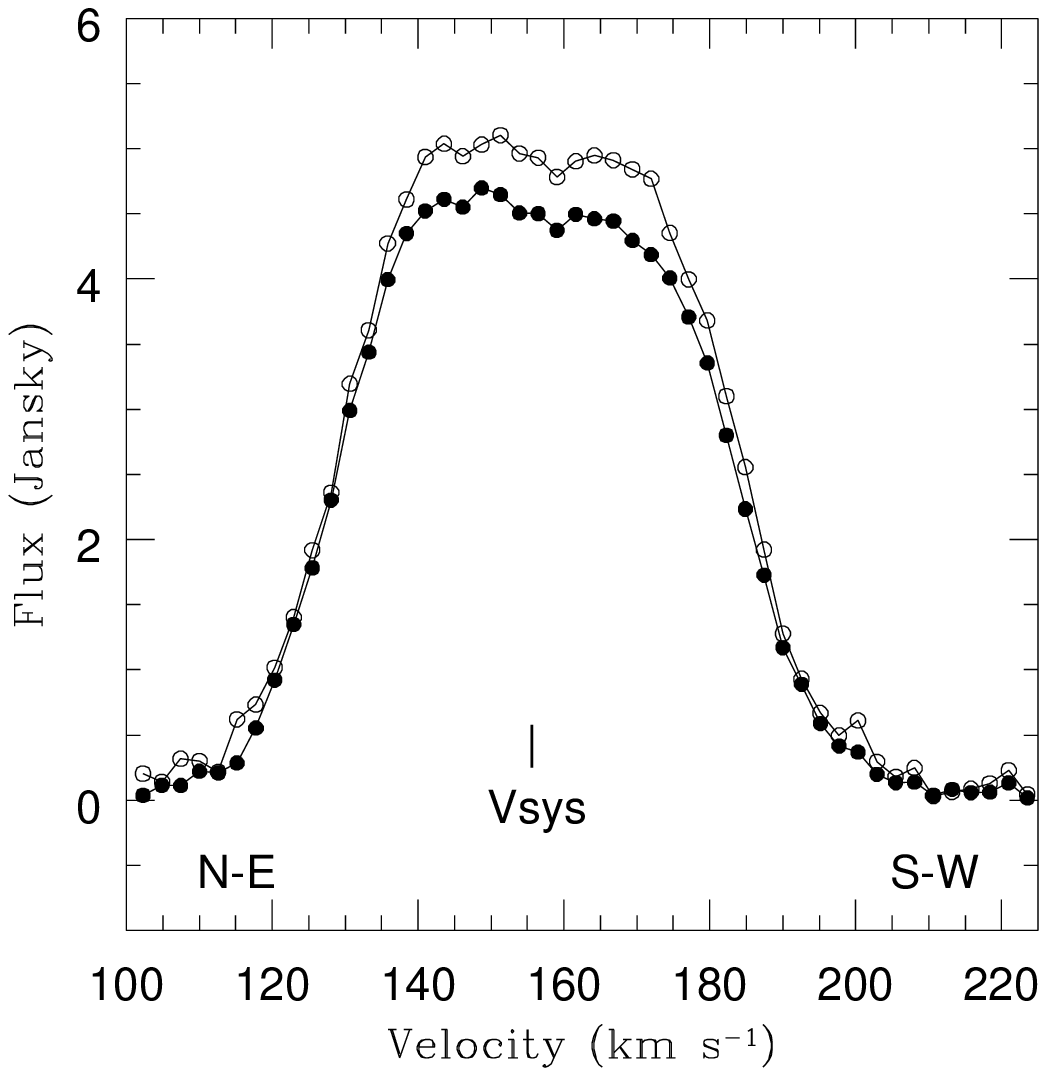]{Global profiles of HoII derived from the
  multi-configuration data. {\em Filled circles:} Profile corrected for the
  non-Gaussian beam shape. {\em Empty circles:} Uncorrected profile.
\label{fig:global_multi}}
%
%
\figcaption[Bureau.fig2_1.ps]{Channel maps of HoII for the D-array data cube. Levels
  are 3, 10, 20, 40, 60, and 80 times 2.5~mJy~beam$^{-1}$
  ($1\sigma=2.75$~mJy~beam$^{-1}$). Positions are B1950 and the heliocentric
  velocity of each channel is indicated in the top-right corner of each map.
  The beam is $66\farcs7\times66\farcs7$, indicated in the bottom-left corner of
  the first map on each page.
\label{fig:channels}}
%
%
\figcaption[Bureau.fig3.ps]{Total \ion{H}{1} map of HoII for the D-array data,
  superposed on a Digitized Sky Survey image. Contours are 0.1, 0.3, 0.6,
  1.0, 2.0, 4.0, 6.0, 10.0, 13.0, 16.0, and 19.0 times $10^{20}$~atoms~cm$^{-2}$ or
  $16.8$~\msun~pc$^{-2}$). The closed contours centered at approximately
  ($\alpha=8\mbox{h} 12\mbox{m} 06\mbox{s}$, $\delta=+70\arcdeg 51\arcmin 05\arcsec$),
  ($\alpha=8\mbox{h} 13\mbox{m} 15\mbox{s}$, $\delta=+70\arcdeg 51\arcmin 20\arcsec$),
  ($\alpha=8\mbox{h} 13\mbox{m} 48\mbox{s}$, $\delta=+70\arcdeg 55\arcmin 05\arcsec$),
  ($\alpha=8\mbox{h} 13\mbox{m} 52\mbox{s}$, $\delta=+70\arcdeg 50\arcmin 45\arcsec$),
  ($\alpha=8\mbox{h} 14\mbox{m} 00\mbox{s}$, $\delta=+70\arcdeg 53\arcmin 45\arcsec$), and
  ($\alpha=8\mbox{h} 14\mbox{m} 25\mbox{s}$, $\delta=+70\arcdeg 55\arcmin 25\arcsec$)
  represent density decrements. The beam is $66\farcs7\times66\farcs7$.
\label{fig:mom0}}
%
%
\figcaption[Bureau.fig4.ps]{\ion{H}{1} velocity field of HoII for the D-array data,
  superposed on a Digitized Sky Survey image. Contours are 120, 130,
  135, 140, 150, 160, 170, 175, and 180~\kms. The closed contours centered at
  approximately 
  ($\alpha=8\mbox{h} 14\mbox{m} 00\mbox{s}$, $\delta=+70\arcdeg 49\arcmin 00\arcsec$) and 
  ($\alpha=8\mbox{h} 13\mbox{m} 50\mbox{s}$, $\delta=+70\arcdeg 57\arcmin 00\arcsec$) 
  have velocities of 135 and 175~\kms\ respectively. The beam is
  $66\farcs7\times66\farcs7$.
\label{fig:mom1}}
%
%
\figcaption[Bureau.fig5.ps]{\ion{H}{1} velocity dispersion field of HoII for the
  D-array data, superposed on a Digitized Sky Survey image. Contours are 6, 9,
  12, 15, 18, and 21~\kms. The beam is $66\farcs7\times66\farcs7$.
\label{fig:mom2}}
%
%
\figcaption[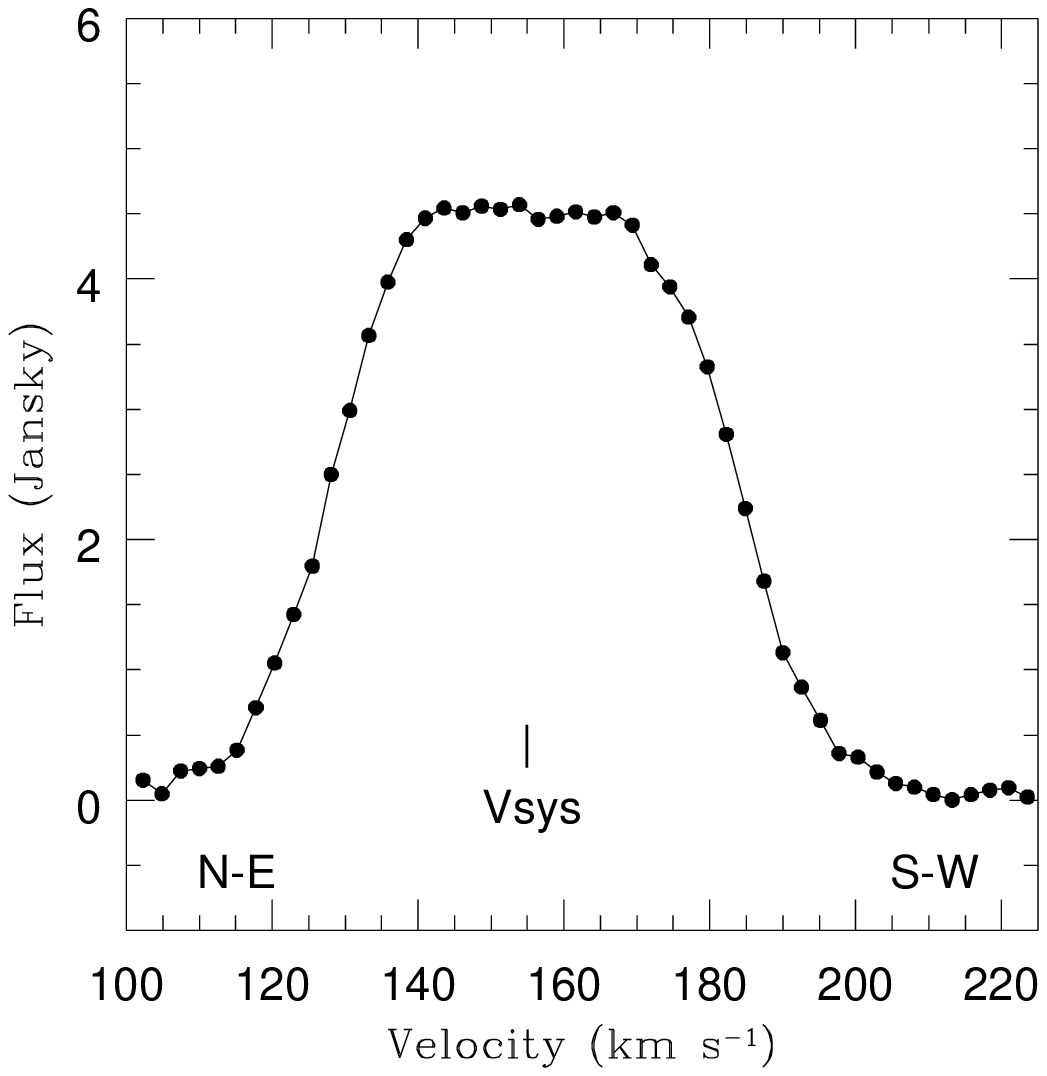]{Global profile of HoII derived from the D-array data.
\label{fig:global_d}}
%
%
\figcaption[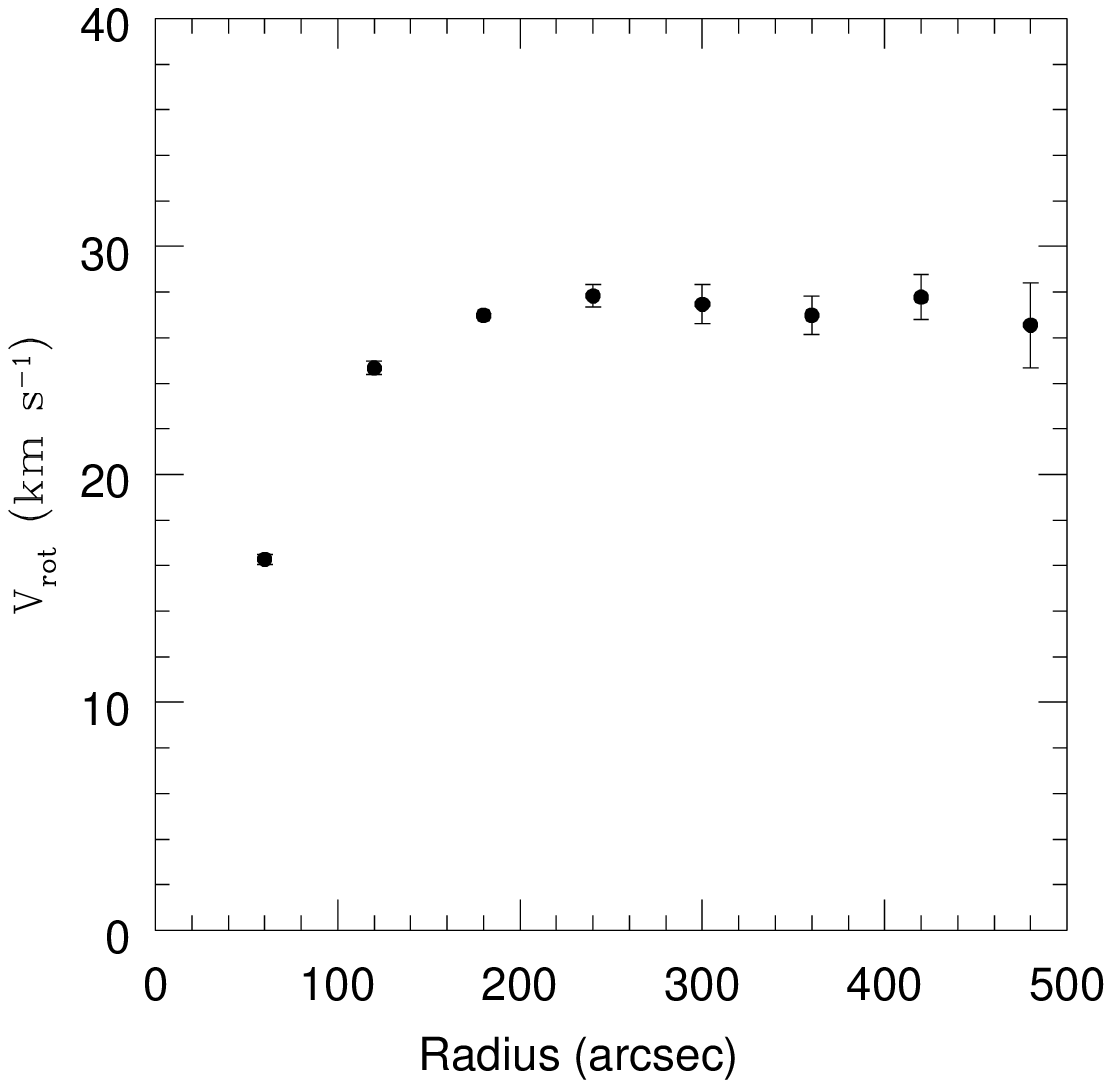]{Rotation curve of HoII for the inner velocity field only
  ($r\lesssim8\arcmin$). The errors represent half the difference between the
  rotation curves of the approaching and receding sides of the galaxy taken
  separately.
\label{fig:rotcur_in}}
%
%
\figcaption[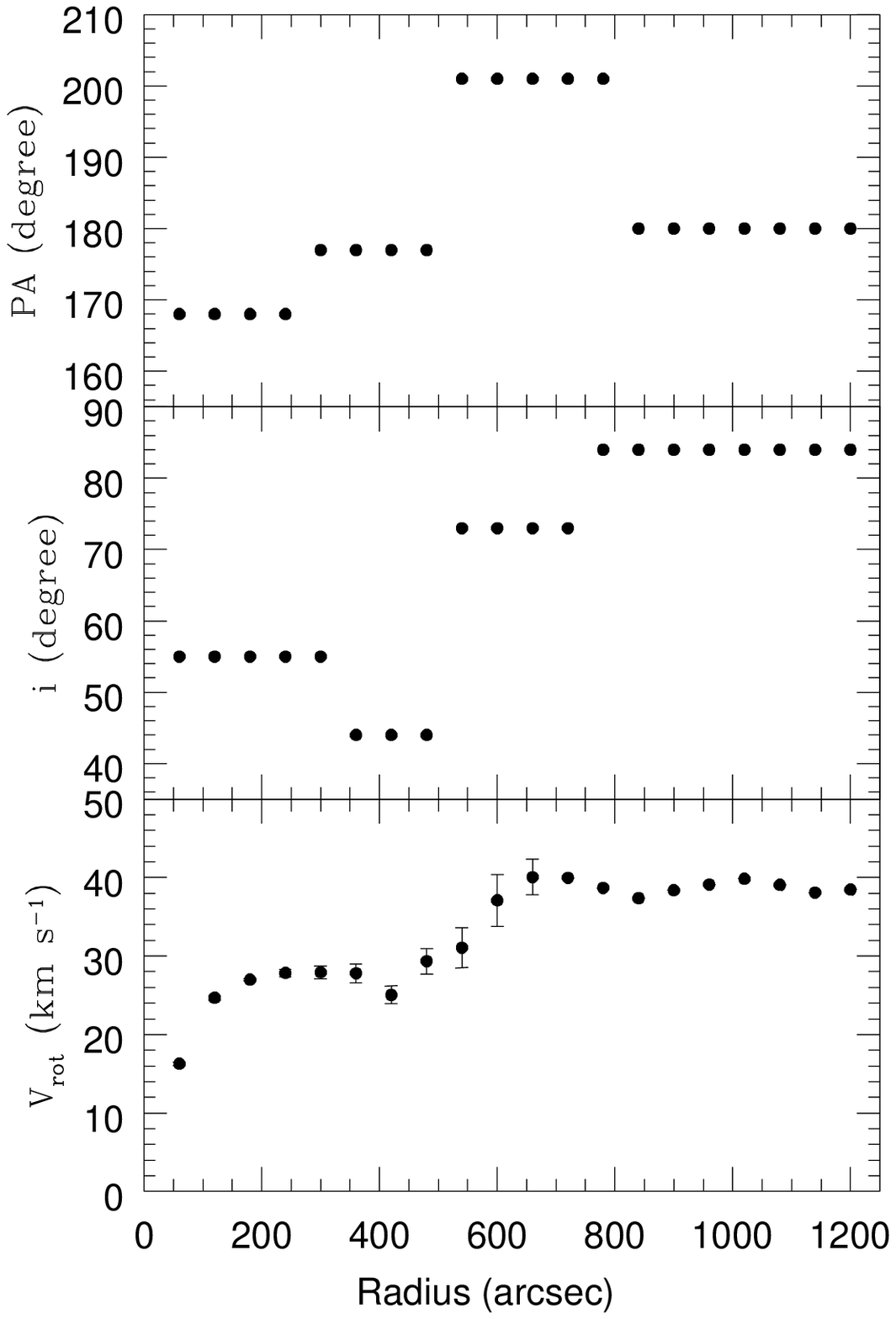]{{\em Top panel:} Adopted variation of PA with radius.
  {\em Middle panel:} Adopted variation of $i$ with radius. {\em Bottom
    panel:} Adopted rotation curve of HoII for the entire velocity field. The
  errors represent half the difference between the rotation curves of the
  approaching and receding sides of the galaxy taken separately. There is no
  error bar past $r=660\arcsec$ as the rotation curve was determined for the
  approaching side only past that point.
\label{fig:rotcur_out}}
%
%
\figcaption[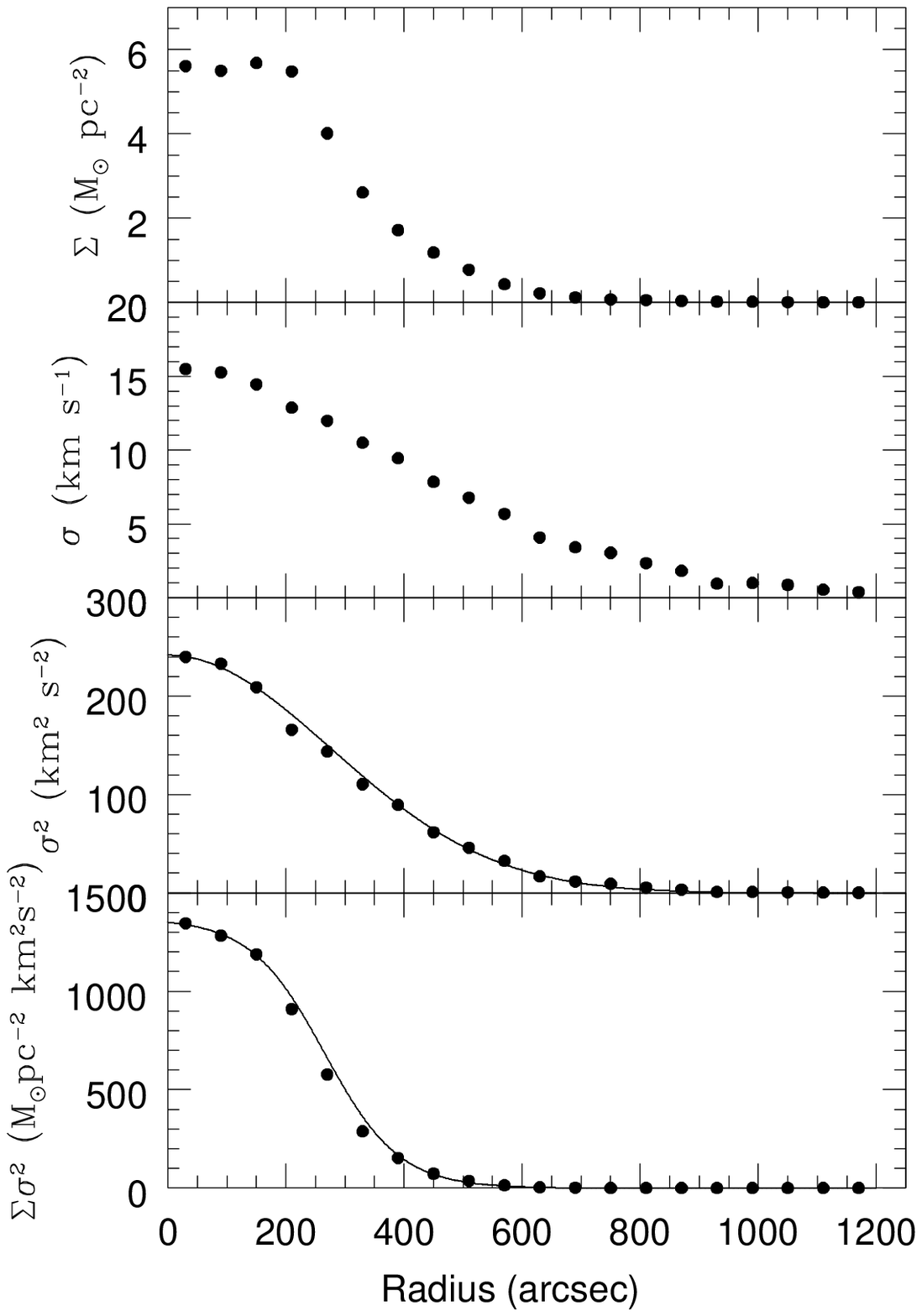]{{\em Top panel:} Radial profile of the deprojected
  \ion{H}{1} surface density $\Sigma$ in HoII. {\em Second panel:} Radial
  profile of the \ion{H}{1} (line-of-sight) velocity dispersion $\sigma$. {\em
    Third panel:} Radial profile of $\sigma^2$. {\em Bottom panel:} Radial
  profile of $\Sigma\sigma^2$. The solid lines are the fits to the data
  discussed in the text (\S~\ref{sec:asym_drift}).
\label{fig:radial_profs}}
%
%
\figcaption[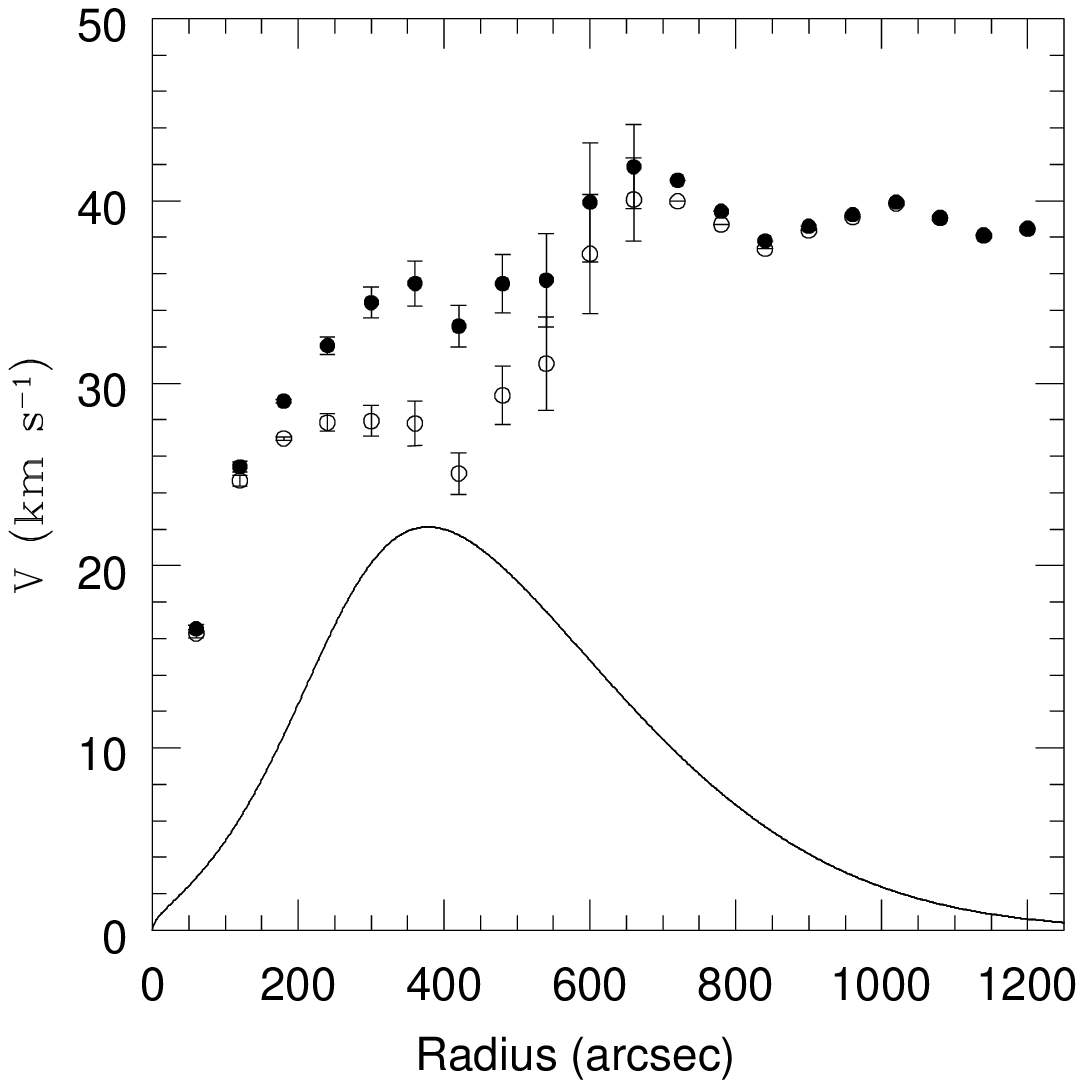]{{\em Filled circles:} Rotation curve of HoII corrected
  for asymmetric drift. {\em Empty circles:} Uncorrected rotation curve. {\em
    Solid curve:} Asymmetric drift correction. The errors represent half the
  difference between the rotation curves of the approaching and receding sides
  of the galaxy taken separately. There is no error bar past $r=660\arcsec$
  because the rotation curve was determined for the approaching side only past
  that point.  No error for the asymmetric drift correction was included.
\label{fig:circ_vel}}
%
%
\figcaption[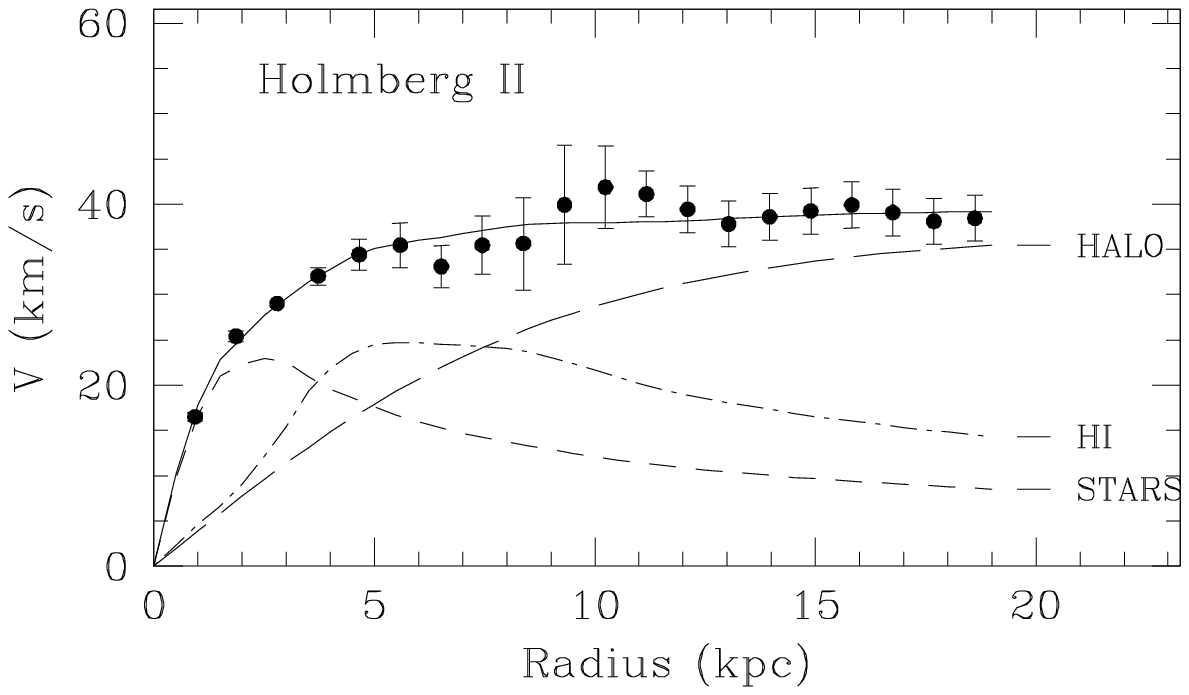]{Mass model of HoII. The contributions of the 3
  components are shown: the stellar disk, \ion{H}{1} disk, and dark isothermal
  halo. In order not to give too much weight to the outer points only defined
  on the approaching side, the mean error from the inner points was used.
\label{fig:mod_mass}}
%
%
%
%
\begin{deluxetable}{llr}
\tablewidth{0pt}
\tablecaption{Basic Properties of HoII \label{ta:basic}}
\tablehead{\colhead{Quantity} & \colhead{Value} & \colhead{Reference}}
\startdata
Right ascension (B1950) & $\alpha=8\mbox{h} 14\mbox{m} 00\fs5$ & 1 \nl
Declination (B1950) & $\delta=+70\arcdeg 52\arcmin 29\farcs8$ & 1 \nl
Heliocentric velocity & $V_{\mbox{\scriptsize \sun}}=156$~\kms & 2 \nl
Distance & $D=3.2$~Mpc & 3 \nl
Morphological type & Im & 4 \nl
Conversion factor & 1$\arcsec=15.5$~pc & \nl
\\
Total apparent $B$ magnitude & $m_{B_T}=11.13$~mag & 1 \nl
Corrected apparent $B$ magnitude & $m^{0,1}_{B_T}=10.51$~mag & 1 \nl
Total absolute $B$ magnitude & $M_{B_T}=-17.01$~mag &  \nl
Apparent radius ($\mu_{B}=25$~mag~arcsec$^{-2}$) & $R_{25}=4\farcm1$ & 1 \nl
Apparent axial ratio ($\mu_{B}=25$~mag~arcsec$^{-2}$) & $b/a=0.87$ & 1 \nl
\enddata
\tablerefs{(1) Lyon/Meudon Extragalactic Database; (2) This work; (3) PWBR92; (4) NASA/IPAC
  Extragalactic Database.}
\end{deluxetable}
\clearpage
%
%
%
\begin{deluxetable}{lr}
\tablewidth{0pt}
\tablecaption{D-Array \ion{H}{1} Observations of HoII \label{ta:h1}}
\tablehead{\colhead{Quantity} & \colhead{Value}}
\startdata
Date of Observations & 1991 March 6 \nl
Total bandwidth & 1.56~MHz \nl
Channel width & 2.58~\kms \nl
Heliocentric central velocity & 190~\kms \nl
Primary beam (HPBW) & $\approx32\arcmin$ \nl
Synthesized beam (FWHM) & $66\farcs7\times46\farcs4$ \nl
rms noise in channel maps & 2.75~mJy~beam$^{-1}$ \nl
Conversion factor (1~mJy~beam$^{-1}$) & 0.14~K \nl
\\
Total \ion{H}{1} Flux & $F_{\mbox{\scriptsize HI}}=267$~Jy~\kms \nl
Total \ion{H}{1} Mass & $M_{\mbox{\scriptsize HI}}=6.44\times10^8$~\msun \nl
\ion{H}{1} Velocity Width & $\Delta V_{50}=57$~\kms \nl
                          & $\Delta V_{20}=72$~\kms \nl
Total Mass (within $r\approx20\arcmin$~=~18.6~kpc) & $M_{\mbox{\scriptsize tot}}=6.3\times10^{9}$~\msun \nl
\enddata
\end{deluxetable}
\clearpage
%
%
%
\begin{deluxetable}{rrrrrrrr}
\tablewidth{0pt}
\tablecaption{HoII D-Array Rotation Curve Parameters \label{ta:rotcur}}
\tablehead{\colhead{\coltwo{$r$}{(arcsec)}} &
           \colhead{\coltwo{PA}{(degree)}} &
           \colhead{\coltwo{$i$}{(degree)}} &
           \colhead{\coltwo{$v_{app}$}{(\kms)}} &
           \colhead{\coltwo{$v_{rec}$}{(\kms)}} &
           \colhead{\coltwo{$v_{rot}$}{(\kms)}} &
           \colhead{\coltwo{$\sigma_D$}{(\kms)}} &
           \colhead{\coltwo{$v_c$}{(\kms)}}}
\startdata
60   & 168 & 55 & 15.82 & 16.71   & 16.28 & 2.87  & 16.53 \nl 
120  & 168 & 55 & 24.10 & 25.26   & 24.67 & 6.14  & 25.42 \nl 
180  & 168 & 55 & 26.80 & 27.16   & 26.98 & 10.68 & 29.02 \nl 
240  & 168 & 55 & 28.80 & 26.88   & 27.84 & 15.90 & 32.06 \nl 
300  & 177 & 55 & 29.55 & 26.17   & 27.93 & 20.14 & 34.43 \nl 
360  & 177 & 44 & 30.27 & 25.37   & 27.80 & 22.03 & 35.47 \nl 
420  & 177 & 44 & 27.32 & 22.70   & 25.05 & 21.68 & 33.13 \nl 
480  & 177 & 44 & 31.23 & 24.81   & 29.34 & 19.92 & 35.46 \nl 
540  & 201 & 73 & 33.81 & 23.58   & 31.08 & 17.48 & 35.66 \nl 
600  & 201 & 73 & 39.25 & 26.13   & 37.09 & 14.80 & 39.93 \nl 
660  & 201 & 73 & 40.64 & 31.46   & 40.08 & 12.14 & 41.88 \nl 
720  & 201 & 73 & 39.99 & \nodata & 39.97 & 9.69  & 41.13 \nl 
780  & 201 & 84 & 38.71 & \nodata & 38.71 & 7.52  & 39.43 \nl 
840  & 180 & 84 & 37.38 & \nodata & 37.38 & 5.68  & 37.81 \nl 
900  & 180 & 84 & 38.38 & \nodata & 38.38 & 4.19  & 38.61 \nl 
960  & 180 & 84 & 39.12 & \nodata & 39.12 & 3.00  & 39.24 \nl 
1020 & 180 & 84 & 39.86 & \nodata & 39.86 & 2.10  & 39.92 \nl 
1080 & 180 & 84 & 39.06 & \nodata & 39.06 & 1.43  & 39.09 \nl 
1140 & 180 & 84 & 38.10 & \nodata & 38.10 & 0.95  & 38.11 \nl 
1200 & 180 & 84 & 38.46 & \nodata & 38.46 & 0.62  & 38.46 \nl 
\enddata
\end{deluxetable}
\clearpage
%
%
%
\begin{figure}
\plotone{Bureau.fig1.ps}
\end{figure}
\clearpage
%
%
%
%
%
%
%
%
%
%
\vspace*{8cm}
\begin{center}
{\Large Figures 2, 3, 4, and 5 available as separate JPEG files only. \\
        See http://www.strw.leidenuniv.nl/$\sim$bureau/pub\_list.html \\
        for full resolution PostScript files.}
\end{center}
\clearpage
%
%
\begin{figure}
\plotone{Bureau.fig6.ps}
\end{figure}
\clearpage
%
%
\begin{figure}
\plotone{Bureau.fig7.ps}
\end{figure}
\clearpage
%
%
\begin{figure}
\plotone{Bureau.fig8.ps}
\end{figure}
\clearpage
%
%
\begin{figure}
\plotone{Bureau.fig9.ps}
\end{figure}
\clearpage
%
%
\begin{figure}
\plotone{Bureau.fig10.ps}
\end{figure}
\clearpage
%
%
\begin{figure}
\plotone{Bureau.fig11.ps}
\end{figure}
\clearpage
\end{document}